\documentclass[10pt,aps,pra,reprint,showpacs,superscriptaddress]{revtex4-1}
\usepackage[english]{babel}
\usepackage[T1]{fontenc}
\usepackage{natbib}
\usepackage{soul}
\usepackage{amssymb}
\usepackage{bm}
\usepackage{amsmath}
\usepackage[mathcal]{eucal}
\usepackage{color}
\usepackage{graphicx}
\usepackage{hyperref}
\usepackage{mathtools}
\usepackage{url}
\usepackage{tikz}
\usetikzlibrary{shapes}
\usetikzlibrary{decorations}
\usetikzlibrary{arrows}
\usetikzlibrary{matrix}

\newcommand {\nc} {\newcommand}
\nc {\beq} {\begin{eqnarray}}
\nc {\eeqn} [1] {\label{#1} \end{eqnarray}}
\nc {\eoln} [1] {\label{#1} \\}
\nc {\eol} {\nonumber \\}
\nc {\rref} [1] {(\ref{#1})}
\nc {\Eq} [1] {Eq.~(\ref{#1})}
\nc {\Ref} [1] {Ref.~\cite{#1}}
\nc {\la} {\mbox{$\langle$}}
\nc {\ra} {\mbox{$\rangle$}}
\nc {\dem} {\mbox{$\frac{1}{2}$}}
\nc {\cP} {\mathcal{P}}
\nc {\cN} {\mathcal{N}}
\nc {\ve} [1] {\mbox{\boldmath $#1$}}
\nc {\arrow} [2] {\mbox{$\mathop{\rightarrow}\limits_{#1 \rightarrow #2}$}}
\nc {\red}[1] {\textcolor{red}{#1}}
\nc {\mc}[3] {\multicolumn{#1}{#2}{#3}}
\nc {\mr}[3] {\multirow{#1}{#2}{#3}}
\nc {\dd}{\, \mathrm{d}}
\nc {\bs}[1]{\boldsymbol{#1}}
\nc {\ket}[1]{\vert #1 \rangle}
\nc {\bra}[1]{\langle #1 \vert}
\nc {\abs}[1]{\vert #1 \vert}
\nc {\avg}[1]{\langle #1 \rangle}
\nc {\braket}[2]{\langle #1 \vphantom{#2} \vert #2 \vphantom{#1} \rangle}
\nc {\abss}[1]{\left| #1 \right|}

\DeclareMathOperator*{\SumInt}{%
\mathchoice%
  {\ooalign{$\displaystyle\sum$\cr\hidewidth$\displaystyle\int$\hidewidth\cr}}
  {\ooalign{\raisebox{.14\height}{\scalebox{.7}{$\textstyle\sum$}}\cr\hidewidth$\textstyle\int$\hidewidth\cr}}
  {\ooalign{\raisebox{.2\height}{\scalebox{.6}{$\scriptstyle\sum$}}\cr$\scriptstyle\int$\cr}}
  {\ooalign{\raisebox{.2\height}{\scalebox{.6}{$\scriptstyle\sum$}}\cr$\scriptstyle\int$\cr}}
}

\begin{document}

\title{Relativistic semiempirical-core-potential calculations in Ca$^+$, Sr$^+$, and Ba$^+$ ions on~Lagrange meshes}
\author{Livio Filippin}
\email[]{Livio.Filippin@ulb.ac.be}
\author{Sacha Schiffmann}
\email[]{Sacha.Schiffmann@ulb.ac.be}
\affiliation{Chimie\;Quantique\;et\;Photophysique,\,C.P.\,160/09,\,Universit\'e\;Libre\;de\;Bruxelles\,(ULB),\,B-1050\,Brussels,\,Belgium}
\author{J\'{e}r\'{e}my Dohet-Eraly}
\email[]{dohet@pi.infn.it}
\affiliation{Istituto Nazionale di Fisica Nucleare, Sezione di Pisa, I-56127 Pisa, Italy}
\author{Daniel Baye}
\email[]{dbaye@ulb.ac.be}
\affiliation{Physique Quantique and Physique Nucl\'{e}aire Th\'{e}orique et Physique Math\'{e}matique, C.P. 229, Universit\'e Libre de Bruxelles (ULB), B-1050 Brussels, Belgium}
\author{Michel Godefroid}
\email[]{mrgodef@ulb.ac.be}
\affiliation{Chimie\;Quantique\;et\;Photophysique,\,C.P.\,160/09,\,Universit\'e\;Libre\;de\;Bruxelles\,(ULB),\,B-1050\,Brussels,\,Belgium}

\date{\today}

\begin{abstract}
Relativistic atomic structure calculations are carried out in alkaline-earth-metal ions using a semiempirical-core-potential approach. The systems are partitioned into frozen-core electrons and an active valence electron. The core orbitals are defined by a Dirac-Hartree-Fock calculation using the \textsc{grasp{\small 2}k} package. The valence electron is described by a Dirac-like Hamiltonian involving a core-polarization potential to simulate the core-valence electron correlation. The associated equation is solved with the Lagrange-mesh method, which is an approximate variational approach having the form of a mesh calculation because of the use of a Gauss quadrature to calculate matrix elements. Properties involving the low-lying metastable $^2D_{3/2,5/2}$ states of Ca$^{+}$, Sr$^{+}$, and Ba$^{+}$ are studied, such as polarizabilities, one- and two-photon decay rates, and lifetimes. Good agreement is found with other theory and observation, which is promising for further applications in alkali-like systems.
\end{abstract}

\pacs{31.15.ap, 03.65.Pm, 32.10.Dk, 02.70.Hm}

\maketitle


\section{Introduction}
\label{sec:intro}

Atomic polarizabilities and forbidden transitions are of much interest due to their various applications, e.g. in optical atomic clocks, which are based on transitions involving long-lived metastable states~\cite{MSC10,LBY15}. Experiments in this field have reached such a high accuracy that relativistic effects are visible and must be precisely accounted for in the calculations~\cite{JJW17,TQS13,SS11,KSA15,KSA17,ASC07,JMC16,Sa10,STD09,JAS09,IS08,ANS12}. Today's most advanced atomic clocks report relative systematic frequency uncertainties below $10^{-17}$~\cite{CHK10,HSL16}. Reaching higher accuracy is limited by small energy shifts resulting from blackbody radiation and quadratic Stark effect~\cite{SS11,Sa10,JAS09}, highly dependent on the accuracy of static and dynamic polarizabilities~\cite{MSC10}.

Singly ionized calcium (Ca$^{+}$), strontium (Sr$^{+}$), and barium (Ba$^{+}$) have been proposed as candidates for optical frequency standards due to the long lifetime of their $^2D_{3/2,5/2}$ states~\cite{LBY15}. Numerous experiments have been performed in these alkaline-earth-metal ions~\cite{KVV95,LAN99,BDL00,KBL05,ABG94,GHL15,BRS99,MLN99,BMN00,GHW87,MS90,LWG05,BEG93,YND97,GBB07,ANH14,NSD86}. Additionally, several theoretical studies have been carried out using many-body approaches~\cite{SSJ17,TQS13,SID06,VGF92,JMC16,JAS09,SID06,IS08,DFG01}. The estimation of these lifetimes involves the study of the competition between the one-photon electric quadrupole ($E2$) and magnetic dipole ($M1$) channels and the two-photon electric dipole ($2E1$) transitions. While the $E2$ and $M1$ decay rates are widely studied, to our knowledge only one prior calculation~\cite{SJS10} of $2E1$ decay rates of the $^2D_{3/2,5/2}$ states has been carried out in these ions.

\textit{Ab initio} methods include electron correlation through explicit electron excitations. The codes based on these approaches enable the calculation of various spectroscopic properties~\cite{Gr07,SJ08,FGB16b}. However, the computational task is important, requiring the diagonalization of large matrices. By contrast, methods introducing a semiempirical core potential simulate the core-valence correlations for an atom with few valence electrons by means of a core-polarization (CP) potential, offering reduced computational times~\cite{MSC10,MZB08,MZ08}. The CP potential is tuned to ensure that the energies of the valence electrons reproduce the observed binding energies. Relativistic semiempirical-core-potential calculations of lifetimes and polarizabilities have been performed in Ca$^+$~\cite{TQS13} and Sr$^+$~\cite{JMC16}, but no such work exists in Ba$^+$.

The present work combines a semiempirical-core-potential Dirac-Hartree-Fock approach (DHFCP) and the Lagrange-mesh method (LMM)~\cite{BH86,Ba15} to study relativistic polarizabilities, one- and two-photon decay rates, and associated lifetimes in Ca$^{+}$, Sr$^{+}$, and Ba$^{+}$ ions. The LMM is an approximate variational approach involving a basis of Lagrange functions related to a set of mesh points associated with a Gauss quadrature \cite{BH86,VMB93,Ba15}. Lagrange functions are continuous functions that vanish at all points of the corresponding mesh but one. The principal simplification appearing in the LMM is that matrix elements are calculated with the Gauss quadrature. The one-body potential matrices are then diagonal and only involve values of the potential at mesh points.

Recently, we have shown that numerically exact solutions of the Dirac-Coulomb equation are obtained with the LMM \cite{Ba15,BFG14}. The method is accurate for most central potentials, such as Yukawa potentials~\cite{BFG14}. It also allows the accurate calculation of polarizabilities and of one- and two-photon decay rates in various types of potentials with small computing times~\cite{FGB14,FGB16}. In this work, the core orbitals obtained from a closed-shell DHF calculation with the \textsc{grasp{\small 2}k} package~\cite{JHF07,JGB13} are projected on Lagrange bases, contrary to Refs.~\cite{TQS13,JMC16} where a DHF program has been developed using $B$ splines or S-spinors. The DHFCP-LMM method is used for single-valence-electron calculations.

In Sec.~\ref{sec:rel_form}, the formulation of the DHFCP method is recalled, and relativistic expressions of polarizabilities and of one- and two-photon decay rates are presented in the case of a single valence electron in a DHFCP potential. In Sec.~\ref{sec:LMM}, the principle of the LMM is summarized and the studied properties are approximated with Gauss quadratures. Section~\ref{sec:res} reports numerical results for low-lying states in Ca$^{+}$, Sr$^{+}$, and Ba$^{+}$, and analyzes the accuracy of the semiempirical-core-potential approach by comparison with \textit{ab initio} calculations and experimental data. Section \ref{sec:conc} contains conclusions.

We use for the fine-structure constant and the atomic unit (a.u.) of time the 2014 CODATA recommended values $1/\alpha=137.035\,999\,139$ and $\hbar/E_h=2.418\,884\,326\,509 \times 10^{-17}$ s~\cite{MNT16}.


\section{Relativistic formulation}
\label{sec:rel_form}

\subsection{Closed-shell DHF equations and core orbitals}
\label{subsec:form_DHF_core}

The starting point of the present approach is a DHF calculation for the closed-shell core state of the atoms. In a.u., the Dirac-Coulomb Hamiltonian for $N_{\text{el}}$ electrons in a central field for a point nucleus of charge $Z$ is given by~\cite{Gr07}
\beq
H_{\text{DC}} = \sum_{i=1}^{N_{\text{el}}} \left[ c \bs{\alpha}_i \cdot \bs{p}_i + (\beta_i - 1)c^2 - \frac{Z}{r_i} \right] + \sum_{i<j}^{N_{\text{el}}} \frac{1}{r_{ij}},
\eeqn{eq_DC_Hamiltonian}
where $c$ is the speed of light and $\bs{\alpha}$ and $\beta$ are the $(4 \times 4)$ Dirac matrices. Since, in the present work, the DHF method is applied to the ground state of closed-shell ions, the total symmetry $J^\Pi$ is equal to $0^+$, with $J$ denoting the total electronic angular momentum and $\Pi$ the parity. For such systems, the total wave function corresponds to a single Configuration State Function (CSF), constructed using anti-symmetrized products of Dirac spinors
\beq
\phi_{n\kappa m} (\ve{r}) = \frac{1}{r} \left( \begin{array}{c} P_{n\kappa}(r) \chi_{\kappa m}(\hat{r}) \\ iQ_{n\kappa}(r) \chi_{-\kappa m}(\hat{r}) \end{array} \right),
\eeqn{eq_phi_nkm}
involving the large and small radial components, $P_{n\kappa}(r)$ and $Q_{n\kappa}(r)$, respectively. The spinor spherical harmonics $\chi_{\kappa m}(\hat{r})$ are common eigenstates of $\ve{L}^2$, $\ve{S}^2$, $\ve{J}^2$, and $J_z$ with respective eigenvalues $l(l+1)$, 3/4, $j(j+1)$, and $m$ where $j = |\kappa| - \dem$ and $l = j + \dem\, \mathrm{sgn}\, \kappa$. The quantum number $n$ labels the different states with the same $\kappa$-symmetry.

The energy of the closed-shell ion is expressed through one-electron integrals $I$ and two-electron Slater integrals $R^k$ as~\cite{Gr07,ZF16}
\beq
& & E_{\text{core}} = \sum_a q_a \, I(a,a) + \sum_a \frac{1}{2} q_a (q_a-1) \left[ R^0(aa,aa) \right. \eol
& & \left. - \frac{[j_a]}{2j_a} \sum_{k=2}^{2l_a} \langle j_a \vert \vert \ve{C}^{(k)} \vert \vert j_a \rangle^2 R^k(aa,aa) \right] + \sum_{a,b>a} q_a q_b \eol
& & \times \left[ R^0(ab,ab) - \sum_{k=\vert l_a-l_b \vert}^{l_a+l_b} \langle j_a \vert \vert \ve{C}^{(k)} \vert \vert j_b \rangle^2 R^k(ab,ba) \right]
\eeqn{eq_E_core}
where the notation $[j]$ means $(2j+1)$, and contributions to the sums over $k$ are not null when $l_a+l_b+k$ is even. Indices $a$ and $b$ refer to one-electron orbitals $n_a\kappa_a$ and $n_b\kappa_b$, respectively, and $q_a$ is the occupation number of orbital~$a$.

The one- and two-electron integrals appearing in \Eq{eq_E_core} are respectively given by
\beq
& & I(a,a) = \int_{0}^{\infty} \left[ -\frac{Z}{r} \, P^2_{a}(r) + c P_{a}(r) \left( -\frac{d}{dr} + \frac{\kappa_{a}}{r} \right) Q_{a}(r) \right. \eol
& & + \left. c Q_{a}(r) \left( \frac{d}{dr} + \frac{\kappa_{a}}{r} \right) P_{a}(r) + \left( -\frac{Z}{r} - 2c^{2} \right) Q^2_{a}(r) \right] dr  \eol
\eeqn{eq_one_electron_I}
and
\beq
& & R^{k}(ab,cd) = \int_{0}^{\infty} \int_{0}^{\infty} \left[ P_{a}(r_{1}) P_{c}(r_{1}) + Q_{a}(r_{1}) Q_{c}(r_{1}) \right] \eol
& & \times \frac{r^{k}_{<}}{r^{k+1}_{>}} \left[ P_{b}(r_{2}) P_{d}(r_{2}) + Q_{b}(r_{2}) Q_{d}(r_{2}) \right] dr_{1}dr_{2},
\eeqn{eq_two_electron_R}
where $r_{<}$ ($r_{>}$) denotes the minimum (maximum) of $r_1$ and $r_2$. The one-electron radial orbitals used to construct the single CSF are determined variationally so as to leave $E_{\text{core}}$, and additional terms for preserving their orthonormality, stationary with respect to their variations. The resulting coupled radial equations are solved iteratively within the self-consistent field procedure, by means of finite difference techniques on an exponential grid~\cite{FGB16b}. The DHF program used in this work is implemented in the \textsc{grasp{\small 2}k} package~\cite{JGB13,JHF07}.

\subsection{DHFCP Hamiltonian and valence orbitals}
\label{subsec:form_DHFCP_valence}

Within the frozen core approximation~\cite{FBJ97,Gr07}, where the relaxation of the core is neglected, the radial DHFCP equation for a single valence electron, denoted by the subscript~$v \equiv n_v\kappa_v$ in the following, is given by
\beq
H_{\text{DHFCP}} \, \phi_{v}(r) = \varepsilon^{\text{DHFCP}}_{v} \, \phi_{v}(r),
\eeqn{eq_eig_DHFCP}
where $\varepsilon^{\text{DHFCP}}_{v}$ is the energy of the valence electron. The Hamiltonian $H_{\text{DHFCP}}$ reads
\beq
\hspace{-0.5cm} H_{\text{DHFCP}} = \begin{pmatrix} -Z/r & c (-\frac{d}{dr} + \frac{\kappa_v}{r}) \\ c (\frac{d}{dr} + \frac{\kappa_v}{r}) & -Z/r - 2c^{2} \end{pmatrix} + V_{\text{core}}(r)
\eeqn{eq_H_DHFCP}
and acts on a 2-component radial wave function $\phi_{v}(r) = (P_{v}(r) \; Q_{v}(r))^T$, where the superscript $T$ means transposition. The semiempirical core potential, $V_{\text{core}}(r)$, appearing in \Eq{eq_H_DHFCP} is defined as the single-electron operator 
\beq
V_{\text{core}}(r) = V_{\text{dir}}(r) + V_{\text{exc}}(r) + V_{\text{CP}}(r),
\eeqn{eq_V_core}
where, in the case of the interaction with a closed-shell core (core electrons are denoted by the subscript~$c \equiv n_c\kappa_c$ in the following), the direct and exchange potentials $V_{\text{dir}}(r)$ and $V_{\text{exc}}(r)$ are defined by their matrix elements~\cite{Gr07,YBF16,JMC16},
\beq
\langle v \vert V_{\text{dir}} \vert v \rangle  &=& \sum_{c \in \text{core}} [j_{c}] \, R^0(vc,vc), \eoln{eq_V_dir}
\langle v \vert V_{\text{exc}} \vert v \rangle &=& -\sum_{c \in \text{core}} \sum_{k} \, [j_{c}] \begin{pmatrix} j_{c} & k & j_{v} \\ 1/2 & 0 & -1/2 \end{pmatrix}^{2} R^k(vc,cv). \eol
\eeqn{eq_V_exc}

The core-polarization potential, $V_{\text{CP}}(r)$, has been introduced to simulate the core-valence correlation neglected in the DHF approximation~\cite{HHJ68,NS76,Hi82,MN88}. The electric field of the valence electron polarizes the core, which acquires an induced dipole moment proportional to the core static dipole polarizability, $\alpha_1(\text{core})$, interacting with the valence electron~\cite{HHJ68,NS76,Hi82,MN88}. Potential $V_{\text{CP}}(r)$ is written as 
\beq
V_{\text{CP}}(r) = -\frac{\alpha_1(\text{core})}{2r^{4}}\left[ 1-\exp{\left( -r^{6}/\rho_\kappa^{6} \right)} \right],
\eeqn{eq_V_CP}
where $\rho_\kappa$ is a cutoff parameter that is tuned to reproduce the experimental binding energy of the lowest state of each $\kappa$-symmetry, and $1-\exp{\left( -r^{6}/\rho_\kappa^{6} \right)}$ is a cutoff function regularizing $V_{\text{CP}}(r)$ at the origin~\cite{TQS13}. Expression~\rref{eq_V_CP} can be extended by taking higher-order corrections into account~\cite{JMC16}.

The present semiempirical approach implies corrections to operators. In particular, when computing matrix elements of 2$^\lambda$-pole transitions between states $n_a\kappa_{a}$ and $n_b\kappa_{b}$, the radial transition operator needs to be modified as~\cite{HHJ68,Ha72,TQS13,JMC16}
\beq
\hspace{-0.5cm} \tilde{r}^\lambda = r^\lambda - \frac{\alpha_\lambda(\text{core})}{r^{\lambda+1}} \sqrt{1 - \exp{[-r^{2(\lambda +2)}/\bar{\rho}^{2(\lambda +2)}}]},
\eeqn{eq_r_tilde_lambda}
where $\alpha_\lambda(\text{core})$ is the static multipole polarizability of the core and $\bar{\rho}$ is the average value $\bar{\rho} = (\rho_{\kappa_{a}}+\rho_{\kappa_{b}})/2$.

\subsection{Polarizabilities}
\label{subsec:form_pola}

The static polarizability of an atomic system can be separated into two terms: a dominant first term from the intermediate valence-excited states, $\alpha(v)$, and a smaller second term from the intermediate core-excited states, $\alpha(\text{core})$~\cite{Sa10}. The latter is smaller than the former by several orders of magnitude~\cite{MSC10}.

For an atomic system described with \Eq{eq_eig_DHFCP}, the dipole polarizability $\alpha_1(n_v\kappa_v m_v)$ of a state $n_v\kappa_v m_v$ with angular momentum $j_v>1/2$ depends on the magnetic projection $m_v$~\cite{MSC10}. It is given by 
\beq
\hspace{-0.25cm} \alpha_{1}(n_v\kappa_v m_v) = \alpha_{1}^{S}(v) + \alpha_{1}^{T}(v) \, \frac{3m_v^2-j_v(j_v+1)}{j_v(2j_v-1)}.
\eeqn{eq_pola}
The quantity $\alpha_{1}^{S}(v)$ is the scalar polarizability while $\alpha_{1}^{T}(v)$ is the tensor polarizability in $j$ representation. The $2^{\lambda}$-pole scalar polarizability reads~\cite{MSC10,FGB14,JMC16}
\beq
& & \alpha^{S}_{\lambda}(v) = \sum_{\kappa'_v} \frac{2[j'_v]}{[\lambda]} \left( \begin{array}{c c c} j'_v & \lambda & j_v \\ -1/2 & 0 & 1/2 \end{array} \right)^2 \eol
& & \times \SumInt_{n'_v} \frac{\{\int_0^\infty [P_{v}(r) P_{v'}(r) + Q_{v}(r) Q_{v'}(r)] \tilde{r}^\lambda dr\}^2}{\varepsilon_{v'}-\varepsilon_{v}}
\eeqn{eq_pola_scalar}
with the subscript $v' \equiv n'_v\kappa'_v$. The radial functions $P_{v}(r)$, $Q_{v}(r)$ and $P_{v'}(r)$, $Q_{v'}(r)$ are solutions of \Eq{eq_eig_DHFCP} with respective energies $\varepsilon_{v}$ and $\varepsilon_{v'}$. The sum over $n'_v$ represents a sum over the discrete states and an integral over the continuum that also involves negative energies. The dipole tensor polarizability is defined as~\cite{MSC10,JMC16}
\beq
& & \alpha^{T}_{1}(v) = 4\, \sqrt{\frac{5j_v(2j_v-1)[j_v]}{6(j_v+1)(2j_v+3)}} \eol
& & \times \sum_{\kappa'_v} (-1)^{j_v+j'_v} [j'_v] \left\lbrace \begin{array}{c c c} j_v & 1 & j'_v \\ 1 & j_v & 2 \end{array} \right\rbrace \left( \begin{array}{c c c} j'_v & 1 & j_v \\ -1/2 & 0 & 1/2 \end{array} \right)^2 \eol
& & \times \SumInt_{n'_v} \frac{\{\int_0^\infty [P_{v}(r) P_{v'}(r) + Q_{v}(r) Q_{v'}(r)] \tilde{r} \, dr\}^2}{\varepsilon_{v'}-\varepsilon_{v}}.
\eeqn{eq_pola_tensor}

\subsection{Decay rates and lifetimes}
\label{subsec:form_decay_rates_lifetimes}

The lifetime (in $s$) of an atomic state is given by the inverse of the sum of all possible decay rates (in $s^{-1}$), $\tau=1/\sum_i W_i$. The dominant one-photon $E2$ and $M1$ and two-photon $2E1$ contributions are studied in this work.

For an atomic system described with \Eq{eq_eig_DHFCP}, the average partial decay rates describing the $2E1$ two-photon transitions reads in a.u.~\cite{GD81,SPI98,FGB16}
\beq
\dfrac{d\overline{W}_{2E1}}{d\omega_1} &=& \dfrac{\omega_1 \omega_2}{8\pi^3c^2[j_i]} \sum_{j_\nu} \left\lbrace \left[ S_{2E1}^{j_\nu}(2,1) \right]^2 + \left[ S_{2E1}^{j_\nu}(1,2) \right]^2 \right. \eol
& & \left. + 2\sum_{j'_\nu} d_{2E1}^{j_\nu,j'_\nu} \, S_{2E1}^{j_\nu}(2,1) S_{2E1}^{j'_\nu}(1,2) \right\rbrace,
\eeqn{eq_dW_2E1}
where the angular coupling factor $d_{2E1}^{j_\nu,j'_\nu}$ is given in \Ref{FGB16}, and $S_{2E1}^{j_\nu}(2,1)$ reads
\beq
S_{2E1}^{j_\nu}(2,1) = \Delta_{2E1}^{j_\nu}(2,1) \sum_{\kappa_\nu}  \SumInt_{n_\nu} \dfrac{\overline{\mathcal{M}}_{f,\nu}^{E1}(\omega_2;G) \, \overline{\mathcal{M}}_{\nu,i}^{E1}(\omega_1;G)}{\varepsilon_{\nu}-\varepsilon_i+\omega_1}. \eol
\eeqn{eq_Sj21}
The angular factor $\Delta_{2E1}^{j_\nu}(2,1)$ is given in \Ref{FGB16}. $S_{2E1}^{j_\nu}(1,2)$ is analogously obtained by permuting indices 1 and~2. Kets $\vert i \rangle \equiv \vert n_i \kappa_i \rangle$ and $\vert f \rangle \equiv \vert n_f \kappa_f \rangle$ correspond to solutions of \Eq{eq_eig_DHFCP} for the initial and final states with respective energies $\varepsilon_i$ and $\varepsilon_f$, and $\omega_j$ is the frequency of the $j$th photon. Energy conservation imposes $\varepsilon_i-\varepsilon_f=\omega_1+\omega_2$, where the recoil of the nucleus is neglected. As for polarizabilities, the transition proceeds through an infinite set of intermediate states $\vert \nu \rangle \equiv \vert n_\nu \kappa_\nu \rangle$ at energy~$\varepsilon_\nu$.

The electric radial matrix elements $\overline{\mathcal{M}}^{EL}$ in \Eq{eq_Sj21} contain an arbitrary gauge parameter~$G$ from which the results should be independent~\cite{Gr74,GD81}. The $G=0$ value defines the Coulomb (or velocity) gauge, which leads to the electric multipole velocity form in the non-relativistic limit. The value $G=\sqrt{(L+1)/L}$ defines the Babushkin (or length) gauge, which leads to the non-relativistic electric multipole length form of the transition operator, and hence allows to account for correction~\rref{eq_r_tilde_lambda}.

In the long-wavelength (LW) approximation~\cite{Gr74,Gr07}, the radial matrix element $\overline{\mathcal{M}}_{\alpha,\beta}^{E1}$ in \Eq{eq_Sj21} reads in the length gauge~\cite{Gr74}
\beq
& & \overline{\mathcal{M}}_{\alpha,\beta}^{E1}(\omega; \sqrt{2}) = \sqrt{2} \left( \frac{\omega}{c} \right) \eol
& & \times \int_0^{\infty} \left[ P_\alpha(r) P_\beta(r) + Q_\alpha(r) Q_\beta(r) \right] \tilde{r} \, dr.
\eeqn{eq_Mfi_E1}

The spontaneous $2E1$ decay rate, $W_{2E1}$, is obtained by integrating $d\overline{W}_{2E1}/d\omega_1$ over $\omega_1$ from 0 to $\varepsilon_i-\varepsilon_f$. The value of $W_{2E1}$ is multiplied by 1/2 to avoid counting twice each pair, because both photons have the same characteristics~\cite{SPI98}.

The spontaneous emission rate for a one-photon transition $i \rightarrow f$ reads in a.u.~\cite{Gr74}
\beq
W_{i \rightarrow f} = \frac{2\omega_t}{c} \, \frac{[j_f]}{[L]} \begin{pmatrix} j_i & L & j_f \\ 1/2 & 0 & -1/2 \end{pmatrix}^2 \vert \overline{\mathcal{M}}_{fi}^{\sigma L}(\omega_t) \vert^2,
\eeqn{eq_one_photon}
where $\sigma=E$ or $M$ and $\omega_t=\varepsilon_i-\varepsilon_f$ is the transition energy. In the LW approximation, $\overline{\mathcal{M}}_{fi}^{E2}$ reads in the length gauge~\cite{Gr74}
\beq
& & \overline{\mathcal{M}}_{fi}^{E2}(\omega_t; \sqrt{3/2}) = \frac{1}{\sqrt{6}} \left( \frac{\omega_t}{c} \right)^2 \eol
& & \times \int_0^{\infty} \left[ P_f(r) P_i(r) + Q_f(r) Q_i(r) \right] \tilde{r}^2 \, dr,
\eeqn{eq_Mfi_E2}
and the gauge-independent radial matrix element $\overline{\mathcal{M}}_{fi}^{M1}$ is given by~\cite{Gr74}
\beq
& & \overline{\mathcal{M}}_{fi}^{M1}(\omega_t) = \frac{1}{\sqrt{2}} \left( \frac{\omega_t}{c} \right) (\kappa_f + \kappa_i) \eol
& & \times \int_0^\infty \left[ P_f(r) Q_i(r) + Q_f(r) P_i(r) \right] r \, dr.
\eeqn{eq_Mfi_M1}
No correction similar to \Eq{eq_r_tilde_lambda} is applied to the magnetic transition operator. The excellent comparison between the present $M1$ decay rates and the reference values~\cite{GJ07} (see \tablename{~\ref{table_lifetimes}}) infers that such corrections would be small.

The use of the LW approximation is justified by the small variation (only on the sixth digit) found in the obtained results when considering $(\omega r/c)^L/(2L+1)!!$ operators instead of spherical Bessel functions $j_L(\omega r/c)$ that occur in the relativistic transition operators~\cite{Gr74}.

The multipole matrix elements involved in the calculation of polarizabilities, one- and two-photon decay rates are dominated by the form of the wave function at long distances from the nucleus. By tuning energies to experimental values, semiempirical-core-potential methods enable to obtain wave functions having the correct asymptotic decrease~\cite{JMC16}.


\section{Lagrange-mesh method}
\label{sec:LMM}

\subsection{Expansions on a Lagrange basis}
\label{subsec:principles}

The principles of the LMM are described in Refs.~\cite{BH86,VMB93,Ba15} and its application to the Dirac equation is presented in Refs.~\cite{Ba15,BFG14}. The mesh points $x_j$ are defined by~\cite{BH86} 
\beq
L_N^{\alpha}(x_j) = 0,
\eeqn{Lag.1}
where $j=1$ to $N$, and $L_N^{\alpha}$ is a generalized Laguerre polynomial~\cite{AS65}. This mesh is associated with a Gauss-Laguerre quadrature 
\beq
\int_0^\infty g(x) \, dx \approx \sum^N_{j=1} \lambda_j \, g(x_j), 
\eeqn{Lag.2}
with the weights $\lambda_j$. Note that the dependence of $x_j$ and $\lambda_j$ on parameter $\alpha$ is implicit. The Gauss quadrature is exact for the Laguerre weight function $x^{\alpha}e^{-x}$ multiplied by any polynomial of degree at most $2N-1$~\cite{Sz67}. The regularized Lagrange functions are defined by~\cite{Ba95,BHV02,Ba15}
\beq
\hat{f}_j^{(\alpha)} (x) = (-1)^j \sqrt{\frac{N !}{\Gamma (N+\alpha+1) x_j}} \; \frac{L_N^{\alpha}(x)}{x-x_j}\, x^{\alpha/2+1} e^{-x/2}. \eol
\eeqn{Lag.3}
The functions $\hat{f}_j^{(\alpha)}(x)$ are polynomials of degree $N-1$ multiplied by $x$ and by the square root of the Laguerre weight $x^{\alpha}e^{-x}$. The Lagrange functions satisfy the Lagrange conditions 
\beq
\hat{f}_j^{(\alpha)}(x_i) = \lambda_i^{-1/2} \delta_{ij}.
\eeqn{Lag.4}
They are not orthonormal, but become orthonormal at the Gauss-quadrature approximation. Condition~\rref{Lag.4} drastically simplifies the expressions of the one-body matrix elements calculated with the Gauss quadrature.

Radial functions $P_{v}(r)$ and $Q_{v}(r)$ are expanded in regularized Lagrange functions~\rref{Lag.3} as 
\beq
P_{v}(r) = h_v^{-1/2} \sum_{j=1}^{N_v} \, p_{v j} \hat{f}_j^{(\alpha)}(r/h),
\eoln{Lag.5}
Q_{v}(r) = h_v^{-1/2} \sum_{j=1}^{N_v} \, q_{v j} \hat{f}_j^{(\alpha)}(r/h),
\eeqn{Lag.6}
where $h_v$ is a scaling parameter aimed at adapting the scaled mesh $\{h_vx_i\}_{i=1}^{N_v}$ to the physical extension of the problem. The parameter $\alpha_v=2(\gamma_v-1)$, where $\gamma_v=\sqrt{\kappa_v^2-(\alpha Z)^2}$, can be selected so that the Lagrange functions behave as $r^{\gamma_v}$ near the origin~\cite{BFG14}. Here, another choice $\alpha_v=2(\gamma_v-|\kappa_v|)$ is preferable~\cite{FGB14,FGB16}. The basis functions then behave as $r^{\gamma_v-|\kappa_v|+1}$, and the physical $r^{\gamma_v}$ behavior can be simulated by linear combinations.

Let us introduce expansions~\rref{Lag.5} and \rref{Lag.6} in \Eq{eq_eig_DHFCP}. Projecting on the Lagrange functions and using the associated Gauss quadrature leads to the $2N_v \times 2N_v$ Hamiltonian matrix
\beq
& & \ve{H}_{\kappa_v}^{\text{DHFCP}} = \ve{H}_{\kappa_v} \eol
& & + \begin{pmatrix} V_{\text{dir}}^{pp} & 0 \\ 0 & V_{\text{dir}}^{qq} \\ \end{pmatrix} + \begin{pmatrix} V_{\text{exc}}^{pp} & V_{\text{exc}}^{pq} \\ V_{\text{exc}}^{qp}  & V_{\text{exc}}^{qq} \\ \end{pmatrix} + \begin{pmatrix} V_{\text{CP}}^{pp} & 0 \\ 0 & V_{\text{CP}}^{qq} \\ \end{pmatrix},
\eeqn{eq_H_DHFCP_mat}
where the $N_v \times N_v$ block matrices verify $V_{\text{dir}}^{pp}=V_{\text{dir}}^{qq}$, $V_{\text{exc}}^{qp} = (V_{\text{exc}}^{pq})^{T}$ and $V_{\text{CP}}^{pp}(i,j)=V_{\text{CP}}^{qq}(i,j)=V_{\text{CP}}(h_v x_i) \, \delta_{ij}$. Matrix $\ve{H}_{\kappa_v}$ corresponds to the $2N_v \times 2N_v$ hydrogen-like Dirac Hamiltonian at the Gauss-quadrature approximation, and reads
\beq
\ve{H}_{\kappa_v} = \left( \begin{array}{c c} -Z/(h_v x_i) \, \delta_{ij} & \frac{c}{h_v} \left( D_{ji}^G + \frac{\kappa_v}{x_i} \delta_{ij} \right) \\ \frac{c}{h_v} \left( D_{ij}^G + \frac{\kappa_v}{x_i} \delta_{ij} \right) & (-Z/(h_v x_i) -2c^2) \, \delta_{ij} \end{array} \right) \eol
\eeqn{Lag.7}
with a $2\times2$ block structure, where the matrix elements $D_{ij} = \langle \hat{f}_i^{(\alpha_v)} \vert d/dx \vert \hat{f}_j^{(\alpha_v)} \rangle$ are calculated at the Gauss-quadrature approximation as
\beq
D_{i \neq j}^G = (-1)^{i-j} \sqrt{\frac{x_i}{x_j}}\, \frac{1}{x_i-x_j}, \quad D_{ii}^G = \frac{1}{2x_i}.
\eeqn{Lag.8}
The diagonalization of Hamiltonian matrix~\rref{eq_H_DHFCP_mat} provides a set of $2N_v$ orthogonal eigenvectors $\ve{p}_v = (p_{v 1},\cdots,p_{v N_v},q_{v 1},\cdots,q_{v N_v})^T$ for each valence orbital $\phi_{v}$ of a given $\kappa_v$-symmetry, and $\sum_{j=1}^{N_v} \left( p_{v j}^2+q_{v j}^2 \right)=1$ ensures the normalization of $P_{v}(r)$ and $Q_{v}(r)$ at the Gauss-quadrature approximation.

\subsection{Evaluation of two-electron Slater integrals}
\label{subsec:evaluation_Rk}

Let us illustrate the calculation of the two-electron Slater integrals $R^k$ by considering e.g. the exchange potential matrix in \Eq{eq_H_DHFCP_mat}. The matrix element $(i,j)$ of the $N_v \times N_v$ block $V_{\text{exc}}^{pq}$ is given by \Eq{eq_V_exc}, where the integral $R^k_{pq}(ic,cj)$ reads, according to \Eq{eq_two_electron_R},
\beq
& & R^k_{pq}(ic,cj) =  h_v^{-1} \int_{0}^{\infty} \int_{0}^{\infty} \hat{f}_i^{(\alpha_v)}(r_{1}/h_v) P_{c}(r_{1}) \eol
& & \times \frac{r^k_{<}}{r^{k+1}_{>}} \, Q_{c}(r_{2}) \hat{f}_j^{(\alpha_v)}(r_{2}/h_v) \, dr_{1} dr_{2}.
\eeqn{eq_Rk_Vexc}

Two integration methods are devised in order to accurately compute $R^k_{pq}(ic,cj)$ with Gauss-Laguerre quadratures. The first one, denoted as ``M~I'', corresponds to the strategy suggested by Hartree~\cite{Ha57,FGB16b}, who first introduced the functions $Y^k(bd; r)$. In the present case~\rref{eq_Rk_Vexc}, $Y_q^k(jc; r)$ reads
\beq
\hspace{-0.5cm} Y_q^k(jc; r) = r \int_0^\infty \frac{r^k_{<}}{r^{k+1}_{>}} \, Q_{c}(s) h_v^{-1/2} \hat{f}_j^{(\alpha_v)}(s/h_v) \, ds
\eeqn{eq_Yk_vc}
with $r \equiv r_1$ and $s \equiv r_2$, and is solution of the second-order differential equation
\beq
& & \frac{d^2}{dr^2} \, Y_q^k(jc; r) - \frac{k(k+1)}{r^2} \, Y_q^k(jc; r) \eol
& & = - \frac{2k+1}{r} \, Q_{c}(r) h_v^{-1/2} \hat{f}_j^{(\alpha_v)}(r/h_v)
\eeqn{eq_diff_Yk}
with the boundary conditions $Y_q^k(jc; 0)=0$ and $dY_q^k(jc; r)/dr \rightarrow -kY_q^k(jc; r)/r+Q_{c}(r) h_v^{-1/2} \hat{f}_j^{(\alpha_v)}(r/h_v)$ as $r \rightarrow \infty$. Special attention is required for the $k=0$ case for which $Y_q^0(jc;\infty) = \int_0^\infty Q_{c}(s) h_v^{-1/2} \hat{f}_j^{(\alpha_v)}(s/h_v) \, ds$, whereas  $Y_q^k(jc;\infty) =0$ for $k>0$. Equation~\rref{eq_diff_Yk} is solved on Lagrange meshes, as presented in the Appendix. Once $Y_q^k(jc; r)$ is known, the two-electron Slater integral~\rref{eq_Rk_Vexc} is expressed as
\beq
R^k_{pq}(ic,cj) = \int_{0}^{\infty} h_v^{-1/2} \hat{f}_i^{(\alpha_v)}(r/h_v) P_{c}(r) \, \frac{1}{r} \, Y_q^k(jc; r) \, dr \eol
\eeqn{eq_Rk_Vexc_2}
and can be computed with the appropriate Gauss-Laguerre quadrature (see the Appendix).

The second method, denoted as ``M~II'', separates $R^k_{pq}(ic,cj)$ in two terms as
\beq
& & R^k_{pq}(ic,cj) =  h_v^{-1} \times \eol
& & \int_{0}^{\infty} \hat{f}^{(\alpha_{v})}_{j}(r_{2}/h_{v}) Q_c(r_2) \, r^{k}_{2} \int_{r_{2}}^{\infty} \frac{\hat{f}^{(\alpha_{v})}_{i}(r_{1}/h_{v}) P_c(r_1)}{r^{k+1}_{1}} dr_{1} dr_{2} \eol
& & + h_v^{-1} \times \eol
& & \int_{0}^{\infty} \frac{\hat{f}^{(\alpha_{v})}_{j}(r_{2}/h_{v}) Q_c(r_2)}{r^{k+1}_{2}} \int_{0}^{r_{2}} \hat{f}^{(\alpha_{v})}_{i}(r_{1}/h_{v}) P_c(r_1) r^{k}_{1} dr_{1} dr_{2} \eol
\eeqn{eq_integral_1}
The first term is denoted as $\mathcal{R}_{pq}^k(ic,cj)$, and the order of integration is permuted in the second term. The integration bounds $r_{1} \in [0,r_{2}]$ and $r_{2} \in [0,\infty[$ become $r_{1} \in [0,\infty[$ and $r_{2} \in [r_{1},\infty[$, leading to
\beq
R^k_{pq}(ic,cj) = \mathcal{R}_{pq}^k(ic,cj) + \mathcal{R}_{qp}^k(jc,ci).
\eeqn{eq_integral_3}	
By using the change of variable $r_{1}=r+r_{2}$, $\mathcal{R}_{pq}^k(ic,cj)$ is integrated over variables $r_{2}$ and $r$, and reads
\beq
& & \mathcal{R}_{pq}^k(ic,cj) = h_v^{-1} \int_{0}^{\infty} \hat{f}^{(\alpha_{v})}_{j}(r_{2}/h_{v}) Q_c(r_2) \, r^{k}_{2} \eol
& & \times \int_{0}^{\infty} \frac{\hat{f}^{(\alpha_{v})}_{i}[(r+r_{2})/h_{v}] P_c(r+r_{2})}{(r+r_{2})^{k+1}} \, dr dr_{2}.
\eeqn{eq_integral_4}
This double integral is evaluated with two different Gauss-Laguerre quadratures, one for each integration variable (see the Appendix). Gauss quadratures of $R^k_{pp}(ic,cj)$ and $R^k_{qq}(ic,cj)$ are analogously obtained using M~I and M~II, as well as the Slater integral $R^0_{pp}(ic,jc)$ involved in the $N_v \times N_v$ block $V_{\text{dir}}^{pp}$ of \Eq{eq_H_DHFCP_mat}.

\subsection{Projection of core orbitals on Lagrange bases}
\label{subsec:projection_core}

Each \textsc{grasp{\small 2}k} core orbital is expanded over a basis of Lagrange functions depending on parameter $\alpha_c=2(\gamma_c-\vert \kappa_c \vert)$. A different value of $h_c$ is assigned to each orbital, and is chosen such that the largest \textsc{grasp{\small 2}k} grid point of orbital~$c$, $R_c$, is at the center of $[h_c x_{N_c-1},h_c x_{N_c}]$, where $x_{N_c-1}$ and $x_{N_c}$ denote the last two Lagrange mesh points, i.e., $h_c = 2R_c/(x_{N_c-1}+x_{N_c})$. To compute the unknown $\{p_{c j}\}_{j=1}^{N_c}$ and $\{q_{c j}\}_{j=1}^{N_c}$ coefficients for each core orbital, the radial wave functions $P_c(r)$ and $Q_c(r)$ are evaluated at each point $h_c x_{i}$ of the scaled Lagrange mesh. Because the \textsc{grasp{\small 2}k} grid points do not correspond to Lagrange mesh points, the wave functions are first interpolated with cubic splines, which are piecewise polynomials, allowing to evaluate the functions at any $r$. Applying Lagrange condition~\rref{Lag.4} respectively yields the expansion coefficients $p_{ci} =  (h_c \lambda_i)^{1/2} \, P_c(h_c x_{i})$ and $q_{ci} =  (h_c \lambda_i)^{1/2} \, Q_c(h_c x_{i})$, for $i=1$ to~$N_c$.

\subsection{Polarizabilities and decay rates on Lagrange meshes}
\label{subsec:pola_photon_mesh}

Polarizabilities and two-photon decay rates proceed through an infinite set of intermediate states with some value of $\kappa'_v$. Finite-basis techniques such as the LMM allow a discretization of the continuum, leading to a truncated sum over $2N'_v$ intermediate states. Some of these states may correspond to approximate eigenstates of \Eq{eq_eig_DHFCP}, while the other ones, discretizing the continuum, have no physical meaning and are called \textit{pseudostates}.

Let $\varepsilon_{n'_v\kappa'_v}$, $n'_v=1,\dots,2N'_v$, be the eigenvalues of matrix $\ve{H}_{\kappa'_v}^{\text{DHFCP}}$ with $\kappa'_v$ replacing $\kappa_v$. The corresponding eigenvectors contain the coefficients $p_{v'j}$ and $q_{v'j}$ of the components $P_{v'}(r)$ and $Q_{v'}(r)$ of the intermediate states. The latter are calculated with $\alpha'_v=2(\gamma'_v-|\kappa'_v|)$ in place of $\alpha_v=2(\gamma_v-|\kappa_v|)$, i.e., matrix $\ve{H}_{\kappa'_v}^{\text{DHFCP}}$ is calculated on a different mesh $h'_vx'_j$ with $N'_v$ mesh points. Hence, the physical intermediate states have the exact behavior $r^{\gamma'_v}$ at the origin. Accurate calculations of Eqs.~\rref{eq_pola_scalar}, \rref{eq_pola_tensor} and \rref{eq_Sj21} with a Gauss-Laguerre quadrature are possible by choosing a third mesh $\bar{h}\bar{x}_i$ where $\bar{h}=2h_vh'_v/(h_v+h'_v)$. In the two-photon case, $h_v$ corresponds to $h_i$ or $h_f$. The $\bar{x}_i$ correspond to the weight function $x^{\bar{\alpha}}e^{-x}$ with $\bar{\alpha}=(\alpha_v+\alpha'_v)/2$, and the corresponding weights are denoted as $\bar{\lambda}_i$.

Approximate scalar polarizabilities $\alpha_{\lambda}^{S}(v)$ are obtained from \Eq{eq_pola_scalar} as
\beq
& & \alpha^{S}_{\lambda}(v) = \sum_{\kappa'_v} \frac{2[j'_v]}{[\lambda]} \left( \begin{array}{c c c} j'_v & \lambda & j_v \\ -1/2 & 0 & 1/2 \end{array} \right)^2 \eol
& & \times \sum_{n'_v=1}^{2N'_v} \frac{\{\int_0^\infty \left[ P_{v}(r) P_{v'}(r) + Q_{v}(r) Q_{v'}(r) \right] \tilde{r}^\lambda dr\}^2}{\varepsilon_{v'}-\varepsilon_{v}}.
\eeqn{Lag.9}
Tensor polarizabilities $\alpha_1^{T}(v)$ are analogously approximated from \Eq{eq_pola_tensor}.

Similarly, approximate $S_{2E1}^{j'}(2,1)$ terms of the $2E1$ decay rates~\rref{eq_dW_2E1} are obtained from \Eq{eq_Sj21} as
\beq
& & S_{2E1}^{j'}(2,1) = \Delta_{2E1}^{j'}(2,1) \eol
& & \times \sum_{\kappa'} \sum_{n'=1}^{2N'} \dfrac{\overline{\mathcal{M}}_{f,n'\kappa'}^{E1}(\omega_2;G) \, \overline{\mathcal{M}}_{n'\kappa',i}^{E1}(\omega_1;G)}{\varepsilon_{n'\kappa'}-\varepsilon_i+\omega_1},
\eeqn{Lag.10}
where the notations $n_\nu$, $j_\nu$ and $\kappa_\nu$ related to the intermediate states $\nu$ are replaced by $n'$, $j'$ and $\kappa'$. $S_{2E1}^{j'}(1,2)$ are analogously approximated. The integral common to Eqs.~\rref{Lag.9} and \rref{Lag.10} is calculated with the Gauss quadrature as
\beq
& & \int_0^\infty \left[ P_{n\kappa}(r) P_{n'\kappa'}(r) + Q_{n\kappa}(r) Q_{n'\kappa'}(r) \right] \tilde{r}^k \, dr \eol
& & \approx \sum_{j=1}^N  \sum_{j'=1}^{N'} \left[ p_{n\kappa j} p_{n'\kappa' j'} + q_{n\kappa j} q_{n'\kappa' j'} \right] \, \mathcal{I}^k_{jj'},
\eeqn{Lag.11}
where $\mathcal{I}^k_{jj'}$ reads
\beq
\mathcal{I}^k_{jj'} &=& \int_0^\infty h^{-1/2} \hat{f}_{j}^{(\alpha)}(r/h) \, \tilde{r}^k \, h'^{-1/2} \hat{f}_{j'}^{(\alpha')}(r/h') \, dr \eol
& \approx & \bar{h} (hh')^{-1/2} \sum_{i=1}^{N_G} \bar{\lambda}_{i} \, \hat{f}_{j}^{(\alpha)}(\bar{h}\bar{x}_i/h) \, \tilde{r}^k(\bar{x}_{i}) \, \hat{f}_{j'}^{(\alpha')}(\bar{h}\bar{x}_i/h'), \eol
\eeqn{Lag.12}
and the non-polynomial factor $\tilde{r}^k(\bar{x}_i)$ is given by
\beq
\tilde{r}^k(\bar{x}_i) &=& (\bar{h} \bar{x}_i)^k \eol
& & - \frac{\alpha_k(\text{core})}{(\bar{h} \bar{x}_i)^{k+1}} \sqrt{1 - \exp{[-(\bar{h} \bar{x}_i)^{2(k+2)}/\bar{\rho}^{2(k+2)}}]}. \eol
\eeqn{Lag.13}
If $\tilde{r}^k$ was replaced by $r^k$, the Gauss quadrature \rref{Lag.2} would be exact with $N_{G} \geq (N+N'+k+1)/2$ mesh points. This suggests to use $N_{G}>(N+N'+k+1)/2$ in the present case. The integrals appearing in the one-photon $E2$ and $M1$ decay rates~\rref{eq_Mfi_E2} and \rref{eq_Mfi_M1} are analogously calculated with a Gauss-Laguerre quadrature.

In the two-photon case, the integral over $\omega_1$ is evaluated with a Gauss-Legendre quadrature using $N_{\omega_1}$ mesh points.


\section{Numerical results}
\label{sec:res}

\subsection{Calculations of core orbitals}
\label{subsec:res_core_orbitals}

The core orbitals of each ion are calculated with \textsc{grasp\small{2}k} and are projected on Lagrange bases. Comparing the original orbitals, the one- and two-electron integrals and the core energies with those computed with the LMM allows to assess the accuracy of the core description within the present approach.

Let us first compute with the LMM the norm, mean values of powers of $r$ and core-orbital overlaps, and compare them with the \textsc{grasp\small{2}k} results. Any orbital from \textsc{grasp\small{2}k} being orthonormal, the error on the norm can be written as $\sqrt{\langle \phi_{n_c\kappa_c} \vert \phi_{n_c\kappa_c} \rangle}-1$, where $\langle \phi_{n_c\kappa_c} \vert \phi_{n_c\kappa_c} \rangle \approx \sum_{j=1}^{N_c} \left( p_{n_c\kappa_c j}^2+q_{n_c\kappa_c j}^2 \right)$ at the Gauss-quadrature approximation. Similarly, mean values of powers of $r$ read $\langle r^{s} \rangle_{n_c\kappa_c} \approx \sum_{j=1}^{N_c} (h_c x_{j})^{s} \left( p_{n_c\kappa_c j}^{2}+q_{n_c\kappa_c j}^{2} \right)$ with the Gauss quadrature, which is exact for $s=-2$ and $-1$. The exactness for $s \geq 0$ is recovered by choosing $N_{G} \geq (2N_c+s+1)/2$ mesh points. The overlap $\langle \phi_{n_{c}\kappa_{c}} \vert \phi_{n'_{c}\kappa_{c}} \rangle$ within the same $\kappa_c$-symmetry is given by Eqs.~\rref{Lag.11} and \rref{Lag.12} with $\kappa=\kappa'=\kappa_c$ and $k=0$. Using a basis of $N_c=50$ Lagrange functions for Ca$^{2+}$, Sr$^{2+}$, and Ba$^{2+}$, all relative errors with respect to \textsc{grasp\small{2}k} are in the range $10^{-7}-10^{-6}$.

Let us now compute with the LMM the core energies~\rref{eq_E_core}, $E_{\text{core}}$, and compare them with the \textsc{grasp\small{2}k} results. Projecting $P_c(r)$ and $Q_c(r)$ on $N_c$ Lagrange functions and using the Gauss quadrature yields $I(c,c) = \ve{p}_c^T \ve{H}_{\kappa_c} \ve{p}_c$ for the one-electron integrals~\rref{eq_one_electron_I}, where $\ve{p}_c = (p_{c 1},\cdots,p_{c N_c},q_{c 1},\cdots,q_{c N_c})^T$, and the $2N_c \times 2N_c$ matrix $\ve{H}_{\kappa_c}$ is given by \Eq{Lag.7} with $\kappa_c$ instead of $\kappa_v$. According to \Eq{eq_E_core}, the total one-electron energy of the core is expressed as $I_{\text{tot}}=\sum_c q_c \, I(c,c)$. The two-electron Slater integrals~\rref{eq_one_electron_I} are computed using methods M~I and M~II. The Appendix provides expressions on Lagrange meshes for the valence case. Similar expressions can be derived for the Slater integrals $R^k(cc,cc)$, $R^0(cc',cc')$ and $R^k(cc',c'c)$. The total two-electron energy of the core, $R_{\text{tot}}$, is expressed as $E_{\text{core}}-I_{\text{tot}}$, i.e., as the sum of the last two terms on the r.h.s. of \Eq{eq_E_core}.

\tablename{~\ref{table_E_core}} displays the Ca$^{2+}$, Sr$^{2+}$, and Ba$^{2+}$ core energies using a basis of $N_c=50$ Lagrange functions. The relative error on $I_{\text{tot}}$ with respect to \textsc{grasp\small{2}k} ranges from $5.7 \times 10^{-7}$ to $7.4 \times 10^{-7}$. Hence, the accuracy on one-electron integrals is of the same order of magnitude as the one on the wave functions themselves. The relative error on $R_{\text{tot}}$ is $2 \times 10^{-6}$, and both M~I and M~II provide the same order of accuracy. Summing the one- and two-electron contributions, the accuracy on $E_{\text{core}}$ is $3 \times 10^{-7}$ for M~I and M~II. Increasing the number of Lagrange functions beyond $N_c=50$ does not improve the accuracy of the results.

\begin{table}[ht!]
\caption{\small{Contribution of one- ($I_{\text{tot}}$) and two-electron ($R_{\text{tot}}$) terms to Ca$^{2+}$, Sr$^{2+}$, and Ba$^{2+}$ core energies, $E_{\text{core}}$ (in a.u.). LMM M~I and M~II values are compared with \textsc{grasp\small{2}k} results. Powers of 10 are indicated within brackets.}}
\vspace{0.1cm}
\begin{tabular}{l c r c r c r}
\hline
\hline
\vspace{0.1cm}
Term                     & \hspace{0.25cm} & \mc{1}{c}{M~I}            & & \mc{1}{c}{M~II}            & & \mc{1}{c}{\textsc{grasp\small{2}k}} \\
\hline
\mc{7}{c}{Ca$^{2+}$} \\
$I_{\text{tot}}$    & \hspace{0.25cm} & \mc{3}{c}{$-$9.197\,475\,6\,[2]}                         & & $-$9.197\,470\,37\,[2]                   \\
$R_{\text{tot}}$   & \hspace{0.25cm} & 2.406\,424\,7\,[2]      & & 2.406\,422\,8\,[2]      & & 2.406\,419\,74\,[2]                        \\
\vspace{0.2cm}
$E_{\text{core}}$ & \hspace{0.25cm} & $-$6.791\,050\,9\,[2] & & $-$6.791\,052\,8\,[2] & & $-$6.791\,050\,63\,[2]                   \\
\mc{7}{c}{Sr$^{2+}$} \\
$I_{\text{tot}}$    & \hspace{0.25cm} & \mc{3}{c}{$-$4.378\,474\,7\,[3]}                         & & $-$4.378\,471\,73\,[3]                   \\
$R_{\text{tot}}$   & \hspace{0.25cm} & 1.200\,919\,8\,[3]      & & 1.200\,919\,5\,[3]      & & 1.200\,917\,63\,[3]                        \\
\vspace{0.2cm}
$E_{\text{core}}$ & \hspace{0.25cm} & $-$3.177\,554\,9\,[3] & & $-$3.177\,555\,2\,[3] & & $-$3.177\,554\,10\,[3]                   \\
\mc{7}{c}{Ba$^{2+}$} \\
$I_{\text{tot}}$    & \hspace{0.25cm} & \mc{3}{c}{$-$1.107\,863\,7\,[4]}                         & & $-$1.107\,862\,86\,[4]                   \\
$R_{\text{tot}}$   & \hspace{0.25cm} & 2.943\,151\,2\,[3]      & & 2.943\,151\,1\,[3]      & & 2.943\,145\,62\,[3]                         \\
\vspace{0.1cm}
$E_{\text{core}}$ & \hspace{0.25cm} & $-$8.135\,485\,8\,[3] & & $-$8.135\,485\,9\,[3] & & $-$8.135\,482\,95\,[3]                    \\
\hline	
\hline
\end{tabular}
\label{table_E_core}
\end{table}

\begin{table}[ht!]
\caption{\small{DHF and DHFCP energies, $\varepsilon_{v}^{\text{DHF}}$ and $\varepsilon_{v}^{\text{DHFCP}}$ (in a.u.), of the five lowest states in Ca$^{+}$, Sr$^{+}$, and Ba$^{+}$. Energies are given relative to the energy of the core. DHF-LMM values are compared with results from \textsc{grasp\small{2}k}.}}
\vspace{0.1cm}
\begin{tabular}{l c c c c c c}
\hline
\hline
\vspace{0.1cm}
                   & \hspace{0.5cm} & \mc{3}{c}{$\varepsilon_{v}^{\text{DHF}}$} & \hspace{0.5cm} & $\varepsilon_{v}^{\text{DHFCP}}$ \\
\cline{3-5}
\vspace{0.1cm}
State           & \hspace{0.5cm} & LMM & \hspace{0.5cm} & \textsc{grasp\small{2}k} & \hspace{0.5cm} & LMM \\
\hline
\multicolumn{7}{c}{Ca$^{+}$} \\
$4s_{1/2}$ & \hspace{0.5cm} & $-$0.416\,626 & \hspace{0.5cm} & $-$0.416\,631\,56 & \hspace{0.5cm} & $-$0.436\,277\,6 \\
$3d_{3/2}$ & \hspace{0.5cm} & $-$0.330\,859 & \hspace{0.5cm} & $-$0.330\,869\,35 & \hspace{0.5cm} & $-$0.374\,082\,8 \\
$3d_{5/2}$ & \hspace{0.5cm} & $-$0.330\,750 & \hspace{0.5cm} & $-$0.330\,759\,53 & \hspace{0.5cm} & $-$0.373\,806\,3 \\
$4p_{1/2}$ & \hspace{0.5cm} & $-$0.309\,994 & \hspace{0.5cm} & $-$0.309\,998\,55 & \hspace{0.5cm} & $-$0.321\,496\,7 \\
\vspace{0.2cm}
$4p_{3/2}$ & \hspace{0.5cm} & $-$0.309\,084 & \hspace{0.5cm} & $-$0.309\,088\,86 & \hspace{0.5cm} & $-$0.320\,481\,1 \\
\multicolumn{7}{c}{Sr$^{+}$} \\
$5s_{1/2}$ & \hspace{0.5cm} & $-$0.382\,915 & \hspace{0.5cm} & $-$0.382\,927\,55 & \hspace{0.5cm} & $-$0.405\,355\,2 \\
$4d_{3/2}$ & \hspace{0.5cm} & $-$0.307\,011 & \hspace{0.5cm} & $-$0.307\,028\,86 & \hspace{0.5cm} & $-$0.339\,033\,6 \\
$4d_{5/2}$ & \hspace{0.5cm} & $-$0.306\,360 & \hspace{0.5cm} & $-$0.306\,378\,05 & \hspace{0.5cm} & $-$0.337\,756\,3 \\
$5p_{1/2}$ & \hspace{0.5cm} & $-$0.284\,816 & \hspace{0.5cm} & $-$0.284\,826\,03 & \hspace{0.5cm} & $-$0.297\,300\,8 \\
\vspace{0.2cm}
$5p_{3/2}$ & \hspace{0.5cm} & $-$0.281\,698 & \hspace{0.5cm} & $-$0.281\,707\,26 & \hspace{0.5cm} & $-$0.293\,649\,1 \\
\multicolumn{7}{c}{Ba$^{+}$} \\
$6s_{1/2}$ & \hspace{0.5cm} & $-$0.343\,264 & \hspace{0.5cm} & $-$0.343\,286\,19 & \hspace{0.5cm} & $-$0.367\,633\,8 \\
$5d_{3/2}$ & \hspace{0.5cm} & $-$0.310\,428 & \hspace{0.5cm} & $-$0.310\,459\,81 & \hspace{0.5cm} & $-$0.345\,426\,9 \\
$5d_{5/2}$ & \hspace{0.5cm} & $-$0.308\,268 & \hspace{0.5cm} & $-$0.308\,299\,53 & \hspace{0.5cm} & $-$0.341\,777\,5 \\
$6p_{1/2}$ & \hspace{0.5cm} & $-$0.260\,904 & \hspace{0.5cm} & $-$0.260\,920\,58 & \hspace{0.5cm} & $-$0.275\,315\,4 \\
\vspace{0.1cm}
$6p_{3/2}$ & \hspace{0.5cm} & $-$0.254\,560 & \hspace{0.5cm} & $-$0.254\,576\,95 & \hspace{0.5cm} & $-$0.267\,611\,3 \\
\hline
\hline			 				 	 
\end{tabular}
 \label{table_DHF_DHFCP}
\end{table}

\subsection{Calculations of valence orbitals}
\label{subsec:res_valence_orbitals}

\tablename{~\ref{table_DHF_DHFCP}} displays DHF and DHFCP energies, $\varepsilon_{v}^{\text{DHF}}$ and $\varepsilon_{v}^{\text{DHFCP}}$ (in a.u.), of the five lowest states in Ca$^{+}$, Sr$^{+}$, and Ba$^{+}$ ions, relative to the core energy. The results are computed with $N_{c}=N_{v}=50$ Lagrange functions. The values of the scaling parameter $h_v$ are 0.10 for $nd_j$ states, 0.11 for $(n+1)s_{1/2}$ states and 0.12 for $(n+1)p_j$ states. In practice, all values between 0.1 and 0.2 are acceptable for these states.

DHF-LMM values, obtained by neglecting $V_{\text{CP}}(r)$ in \Eq{eq_V_core}, are compared with results from \textsc{grasp\small{2}k} at the frozen-core approximation. The relative error on $\varepsilon_{v}^{\text{DHF}}$ with respect to \textsc{grasp\small{2}k} is similar for M~I and M~II, and slightly increases with $Z$, ranging from $1.4 \times 10^{-5}$ to $3.3 \times 10^{-5}$ in Ca$^{+}$, from $3.3 \times 10^{-5}$ to $5.9 \times 10^{-5}$ in Sr$^{+}$, and from $6.3 \times 10^{-5}$ to $9.9 \times 10^{-5}$ in Ba$^{+}$. Besides, the fine-structure splittings are well reproduced for the $nd_j$ and $(n+1)p_j$ states. The accuracy of the present valence calculations is sufficient to obtain reliable results for polarizabilities, one- and two- photon decay rates and associated lifetimes when adding the contribution of $V_{\text{CP}}(r)$, as shown in Secs.~\ref{subsec:res_pola} and \ref{subsec:res_2E1_lifetimes}. Increasing the number of Lagrange functions beyond $N_v=50$ does not improve the accuracy of the results.

The values of the core static dipole polarizabilities used in the DHFCP-LMM calculations are computed at the relativistic random-phase approximation (RRPA) and are taken from \Ref{JKH83}: $\alpha_1(\text{Ca}^{2+})=3.254$ a.u., $\alpha_1(\text{Sr}^{2+})=5.813$ a.u., and $\alpha_1(\text{Ba}^{2+})=10.61$ a.u. The $\rho_{\kappa_v}$-values (in a.u.), listed in \tablename{~\ref{table_rho_Kv}}, ensure that the relative error on $\varepsilon_{v}^{\text{DHFCP}}$ with respect to experimental NIST data~\cite{KRR15} is below $10^{-7}$ for all states presented in \tablename{~\ref{table_DHF_DHFCP}}, considering both M~I and M~II. The obtained $\rho_{\kappa_v}$-numbers show good agreement with the ones provided by \Ref{TQS13} for Ca$^{+}$ and by \Ref{JMC16} for Sr$^{+}$.

\begin{table}[ht!]
\caption{\small{Cutoff parameters, $\rho_{\kappa_v}$ (in a.u.), for different $\kappa_v$-symmetries in Ca$^{+}$, Sr$^{+}$, and Ba$^{+}$. Comparison with Refs.~\cite{TQS13,JMC16}.}}
\vspace{0.1cm}
\begin{tabular}{l c c c c c c c c c c}
\hline
\hline
\vspace{0.1cm}
                   & \hspace{0.25cm} & \mc{9}{c}{$\rho_{\kappa_v}$ (a.u.)} \\
\cline{3-11}
\vspace{0.1cm}
                   & \hspace{0.25cm} & \mc{3}{c}{Ca$^{+}$}                                    & \hspace{0.25cm} & \mc{3}{c}{Sr$^{+}$}                                  & \hspace{0.25cm} & Ba$^{+}$ \\
\cline{3-5} \cline{7-9}
\vspace{0.1cm}
$\kappa_v$ & \hspace{0.25cm} & LMM        & \hspace{0.25cm} & \Ref{TQS13} & \hspace{0.25cm} & LMM       & \hspace{0.25cm} & \Ref{JMC16} & \hspace{0.25cm} & LMM        \\
\hline
$-1$            & \hspace{0.25cm} & 1.73808 & \hspace{0.25cm} & 1.7419         & \hspace{0.25cm} & 2.02900 & \hspace{0.25cm} & 2.04960       & \hspace{0.25cm} & 2.35081 \\
$+1$           & \hspace{0.25cm} & 1.63549 & \hspace{0.25cm} & 1.6389         & \hspace{0.25cm} & 1.94914 & \hspace{0.25cm} & 1.97169       & \hspace{0.25cm} & 2.24066 \\	
$-2$            & \hspace{0.25cm} & 1.63216 & \hspace{0.25cm} & 1.6354         & \hspace{0.25cm} & 1.95229 & \hspace{0.25cm} & 1.97600       & \hspace{0.25cm} & 2.26242 \\	
$+2$           & \hspace{0.25cm} & 1.84605 & \hspace{0.25cm} & 1.8472         & \hspace{0.25cm} & 2.34998 & \hspace{0.25cm} & 2.35353       & \hspace{0.25cm} & 2.75043 \\
\vspace{0.1cm}
$-3$            & \hspace{0.25cm} & 1.84776 & \hspace{0.25cm} & 1.8489         & \hspace{0.25cm} & 2.36151 & \hspace{0.25cm} & 2.36534       & \hspace{0.25cm} & 2.77960 \\
\hline
\hline
\end{tabular}
\label{table_rho_Kv}	
\end{table}

\subsection{Calculations of polarizabilities}
\label{subsec:res_pola}

\tablename{s~\ref{table_pola_dipole}} and~\ref{table_pola_quadrupole} respectively display static scalar dipole $(\alpha_1^{S})$ and quadrupole $(\alpha_2^{S})$ polarizabilities (in a.u.) of the five lowest states in Ca$^+$, Sr$^+$, and Ba$^+$ ions. Tensor dipole $(\alpha_1^{T})$ polarizabilities (in a.u.) are given for the $j_v>1/2$ states. Core dipole and quadrupole polarizabilities from \Ref{JKH83} are added to the valence scalar values, $\alpha_1^{S}(v)$ and $\alpha_2^{S}(v)$. The core quadrupole values are $\alpha_2(\text{Ca}^{2+})=6.936$~a.u., $\alpha_2(\text{Sr}^{2+})=17.15$~a.u., and $\alpha_2(\text{Ba}^{2+})=45.96$~a.u.

The DHFCP-LMM results are computed with $N_c=N_v=N'=50$ Lagrange functions, where $N'$ denotes the number of functions used to describe the intermediate states. Significant digits are estimated by increasing $N'$ from 50 to 80. The LMM calculations are performed with M~I and M~II, and the comparison of their results allows to assess the precision of the values displayed in \tablename{s~\ref{table_pola_dipole}} and~\ref{table_pola_quadrupole}.

A more stringent estimate of the precision achieved by the DHFCP-LMM approach is given by studying the effect of variations in the values of the core dipole and quadrupole polarizabilities on the final results. A second set of core dipole values is provided by \Ref{CMA13}: $\alpha_1(\text{Ca}^{2+})=3.284$~a.u., $\alpha_1(\text{Sr}^{2+})=5.748$~a.u., and $\alpha_1(\text{Ba}^{2+})=10.426$~a.u. The relative differences with respect to the first set of values are respectively $0.9\%$, $1.1\%$ and $1.7\%$. A second set of core quadrupole values is provided by \Ref{ANS12} for Ca$^{2+}$ and Sr$^{2+}$, and by \Ref{IS08} for Ba$^{2+}$: $\alpha_2(\text{Ca}^{2+})=6.15$~a.u., $\alpha_2(\text{Sr}^{2+})=14.50$~a.u., and $\alpha_2(\text{Ba}^{2+})=44$~a.u. The relative differences with respect to the first set of values are respectively $11.3\%$, $15.5\%$ and $4.3\%$, thus substantially higher than for the core dipole values.

Using the second set of core polarizabilities in Eqs.~\rref{eq_V_CP} and \rref{eq_r_tilde_lambda} enables to estimate theoretical uncertainties on the values displayed in \tablename{s~\ref{table_pola_dipole}} and~\ref{table_pola_quadrupole}. Note that using another set of core dipole values implies to determine other $\rho_{\kappa_v}$-values in \Eq{eq_V_CP}, to ensure relative errors on $\varepsilon_{v}^{\text{DHFCP}}$ below $10^{-7}$ with respect to NIST data.

In order to study the accuracy of the DHFCP-LMM approach, the present results are compared with other semiempirical-core-potential approaches (RCICP--relativistic configuration interaction with a semiempirical core potential, DFCP--Dirac-Fock with a semiempirical core potential), with \textit{ab initio} methods (RMBPT-SD--relativistic many-body perturbation theory with single and double contributions, R(L)CCSD(T)--relativistic (linearized) coupled cluster method with single, double (and partial triple) contributions), and with experimental works (SA--spectral analysis, DFI--delayed field ionization, RESIS--resonant excitation Stark ionization spectroscopy).

\vspace{-0.25cm}

\subsubsection{Dipole polarizabilities}
\label{subsubsec:res_pola_dipole}

The ground $(n+1)s_{1/2}$ dipole polarizabilities involve $n'p_{1/2,3/2}$ intermediate states. They are dominated by the resonant $(n+1)s_{1/2} \rightarrow (n+1)p_j$ transitions, and their accuracy is largely dependent on the accuracy of the transition matrix elements connecting these states. Excellent consistency is found with RCICP~\cite{JJW17} and DFCP~\cite{TQS13} for Ca$^+$ ($n=3$) and with RCICP~\cite{JMC16} for Sr$^+$ ($n=4$), while no reference value is available with these approaches for Ba$^+$ ($n=5$). The agreement with \textit{ab initio} calculations is satisfactory, although the present method tends to underestimate $\alpha_1^{S}$ by a few percents. This is also true for the RCICP and DFCP methods, and is a direct consequence of the slightly different line strengths for the resonant transitions in these two types of calculations. Good agreement with observation is found for Ca$^+$, while the large uncertainty of the experimental value for Sr$^+$ cannot be used to discriminate between theoretical estimates. By contrast, \textit{ab initio} calculations are more consistent with experiment than the present one for Ba$^+$.

The $nd_j$ dipole polarizabilities involve $n'p_{1/2,3/2},n'f_{5/2}$ states for $nd_{3/2}$, and $n'p_{3/2},n'f_{5/2,7/2}$ states for $nd_{5/2}$. The results of $\alpha_1^{S}$ and $\alpha_1^{T}$ are consistent with RCICP and DFCP for Ca$^+$ and Sr$^+$. The agreement with \textit{ab initio} methods is satisfactory for Ca$^+$ and Sr$^+$, while for Ba$^+$ better agreement is found with \Ref{STD09} than with Refs.~\cite{KSA15,KSA17}.

The $(n+1)p_j$ dipole polarizabilities involve $n's_{1/2},n'd_{3/2}$ states, and additional $n'd_{5/2}$ states for $(n+1)p_{3/2}$. Negative $\alpha_1^{S}$ values for Ca$^+$ and Sr$^+$ arise from negative oscillator strengths of the transitions to $(n+1)s_{1/2}$ and $nd_j$. For $4p_j$ states in Ca$^+$, cancellations in the sum lead to small $\alpha_1^{S}$ values, and consistency with RCICP and DFCP is poor. Moreover, the \textit{ab initio} works do not agree with each other. Better agreement is found between $\alpha_1^{T}$ values of $4p_{3/2}$. For $5p_j$ states in Sr$^+$, excellent consistency is obtained with RCICP. Results are in good agreement with \textit{ab initio} methods for $\alpha_1^{S}$, while the present $\alpha_1^{T}$ value of $5p_{3/2}$ is around $8\%$ smaller because the matrix element of $5s_{1/2} \rightarrow 5p_{3/2}$ is smaller. Only one \textit{ab initio} calculation is available for $6p_j$ states in Ba$^+$. The agreement is poor ($10\%$ difference) for $6p_{1/2}$. For $6p_{3/2}$, $\alpha_1^{S}$ values agree well, while $\alpha_1^{T}$ results disagree because of oscillator strengths cancellations, associated with a higher uncertainty.

\onecolumngrid

\begin{table}[ht!]
\caption{\small{Static scalar dipole $(\alpha_1^{S})$ polarizabilities (in a.u.) of the five lowest states in Ca$^+$, Sr$^+$, and Ba$^+$. Tensor dipole $(\alpha_1^{T})$ polarizabilities (in a.u.) are given for the $j_v>1/2$ states. Core polarizabilities from \Ref{JKH83} are added to the DHFCP-LMM valence $\alpha_1^{S}(v)$ results. Comparison with other theory and experiment. Uncertainties in the last digits are given within parentheses.}}
\vspace{0.1cm}
\resizebox{\textwidth}{!}{
\begin{tabular}{l c l c l l c l l c l c l l}
\hline
\hline
\mc{14}{c}{Ca$^+$} \\
\vspace{0.1cm}
             & \hspace{0.25cm} & \mc{1}{c}{$4s_{1/2}$}                  & \hspace{0.25cm} & \mc{2}{c}{$3d_{3/2}$}                                                    & \hspace{0.25cm} & \mc{2}{c}{$3d_{5/2}$}                                                     & \hspace{0.25cm} & \mc{1}{c}{$4p_{1/2}$}      & \hspace{0.25cm} & \mc{2}{c}{$4p_{3/2}$} \\
\cline{5-6} \cline{8-9} \cline{13-14}
\vspace{0.1cm}
Method  & \hspace{0.25cm} & \mc{1}{c}{$\alpha_1^{S}$}            & \hspace{0.25cm} & \mc{1}{c}{$\alpha_1^{S}$} & \mc{1}{c}{$\alpha_1^{T}$} & \hspace{0.25cm} & \mc{1}{c}{$\alpha_1^{S}$} & \mc{1}{c}{$\alpha_1^{T}$} & \hspace{0.25cm} & \mc{1}{c}{$\alpha_1^{S}$} & \hspace{0.25cm} & \mc{1}{c}{$\alpha_1^{S}$} & \mc{1}{c}{$\alpha_1^{T}$} \\
\hline
DHFCP-LMM                     & \hspace{0.25cm} & 75.272(24)          & \hspace{0.25cm} & 32.986(10)                            & $-$17.884(18)                 & \hspace{0.25cm} & 32.814(10)                         & $-$25.174(26)                      & \hspace{0.25cm} & $-$3.408(78)                    & \hspace{0.25cm} & $-$1.584(77)        & 10.202(20) \\
RCICP~\cite{JJW17}        & \hspace{0.25cm} & 75.46(72)             & \hspace{0.25cm} & 32.98(24)                             & $-$17.97(17)                   & \hspace{0.25cm} & 32.80(24)                           & $-$25.28(24)                         & \hspace{0.25cm} & $-$2.98(11)                     & \hspace{0.25cm} & $-$1.12(10)           & 10.20(11)  \\
DFCP~\cite{TQS13}         & \hspace{0.25cm} & 75.28                   & \hspace{0.25cm} & 32.99                                    & $-$17.88                          & \hspace{0.25cm} & 32.81                                 & $-$25.16                               & \hspace{0.25cm} & $-$2.774                           & \hspace{0.25cm} & $-$0.931                & 10.12        \\
RMBPT-SD~\cite{SS11}    & \hspace{0.25cm} & 76.1(5)                & \hspace{0.25cm} & 32.0(3)                                  & $-$17.43(23)                   & \hspace{0.25cm} & 31.8(3)                              & $-$24.51(29)                        & \hspace{0.25cm} & $-$0.75(70)                       & \hspace{0.25cm} & \;\, 1.02(64)          & 10.31(28)  \\
RCCSD~\cite{KSA15}       & \hspace{0.25cm} & 76.03                   & \hspace{0.25cm} & 32.3                                      & $-$17.02                          & \hspace{0.25cm} & 32.05                                 & $-$23.92                              & \hspace{0.25cm} & \;\, 0.82                             & \hspace{0.25cm} & \;\, 2.82                 & 10.08        \\
RCCSD(T)~\cite{KSA17}  & \hspace{0.25cm} & 76.1(2)                 & \hspace{0.25cm} & 33.67(180)                           & $-$17.71                          & \hspace{0.25cm} & 33.11(180)                        & $-$24.78(4)                          & \hspace{0.25cm} &                                           & \hspace{0.25cm} &                               &                  \\
RLCCSD(T)~\cite{ASC07} & \hspace{0.25cm} & 76.1(11)               & \hspace{0.25cm} &                                             &                                         & \hspace{0.25cm} & 32.0(11)                            &                                              & \hspace{0.25cm} &                                           & \hspace{0.25cm} &                               &                  \\
Expt. SA~\cite{Ch83}       & \hspace{0.25cm} & 75.3(4)                 & \hspace{0.25cm} &                                             &                                         & \hspace{0.25cm} &                                           &                                             & \hspace{0.25cm} &                                           & \hspace{0.25cm} &                               &                  \\
\mc{14}{c}{Sr$^+$} \\
\vspace{0.1cm}
            & \hspace{0.25cm} & \mc{1}{c}{$5s_{1/2}$}                   & \hspace{0.25cm} & \mc{2}{c}{$4d_{3/2}$}                                                    & \hspace{0.25cm} & \mc{2}{c}{$4d_{5/2}$}                                                     & \hspace{0.25cm} & \mc{1}{c}{$5p_{1/2}$}       & \hspace{0.25cm} & \mc{2}{c}{$5p_{3/2}$} \\
\cline{5-6} \cline{8-9} \cline{13-14}
\vspace{0.1cm}
& \hspace{0.25cm} & \mc{1}{c}{$\alpha_1^{S}$} & \hspace{0.25cm} & \mc{1}{c}{$\alpha_1^{S}$} & \mc{1}{c}{$\alpha_1^{T}$} & \hspace{0.25cm} & \mc{1}{c}{$\alpha_1^{S}$} & \mc{1}{c}{$\alpha_1^{T}$} & \hspace{0.25cm} & \mc{1}{c}{$\alpha_1^{S}$} & \hspace{0.25cm} & \mc{1}{c}{$\alpha_1^{S}$} & \mc{1}{c}{$\alpha_1^{T}$}\\
\hline
DHFCP-LMM                    & \hspace{0.25cm} & 89.708(36)           & \hspace{0.25cm} & 63.102(14)                           & $-$35.072(58)                  & \hspace{0.25cm} & 61.979(10)                          & $-$47.325(76)                     & \hspace{0.25cm} & $-$31.69(21)                      & \hspace{0.25cm} & $-$21.43(20)        & 9.802(41)   \\
RCICP~\cite{JMC16}       & \hspace{0.25cm} & 90.10(127)            & \hspace{0.25cm} & 63.12(82)                             & $-$35.11(50)                   & \hspace{0.25cm} & 61.99(72)                             & $-$47.38(67)                       & \hspace{0.25cm} & $-$31.29(49)                     & \hspace{0.25cm} & $-$20.92(70)        & 9.836(147) \\
RMBPT-SD~\cite{Sa10}   & \hspace{0.25cm} & 92.2(7)                 & \hspace{0.25cm} & 63.3(9)                                 & $-$35.5(6)                        & \hspace{0.25cm} & 62.0(9)                                 & $-$47.7(8)                           & \hspace{0.25cm} & $-$32.2(9)                         & \hspace{0.25cm} & $-$21.4(8)            & 10.74(23)   \\
RCCSD~\cite{KSA15}      & \hspace{0.25cm} & 90.54                   & \hspace{0.25cm} & 63.74                                    & $-$35.26                           & \hspace{0.25cm} & 62.08                                   & $-$47.35                              & \hspace{0.25cm} & $-$31.27                            & \hspace{0.25cm} & $-$20.79               & 10.52         \\
RCCSD(T)~\cite{KSA17}  & \hspace{0.25cm} & 91.23(30)            & \hspace{0.25cm} & 64.7(25)                               & $-$35.88(5)                      & \hspace{0.25cm} & 63.5(25)                               & $-$48.29(7)                         & \hspace{0.25cm} &                                           & \hspace{0.25cm} &                              &                   \\
RCCSD(T)~\cite{STD09}  & \hspace{0.25cm} & 88.29(100)          & \hspace{0.25cm} & 61.43(52)                             & $-$35.42(25)                    & \hspace{0.25cm} & 62.87(75)                             & $-$48.83(30)                       & \hspace{0.25cm} &                                           & \hspace{0.25cm} &                              &                   \\
RLCCSD(T)~\cite{JAS09} & \hspace{0.25cm} & 91.3(9)               & \hspace{0.25cm} &                                              &                                          & \hspace{0.25cm} & 62.0(5)                                 &                                             & \hspace{0.25cm} &                                           & \hspace{0.25cm} &                              &                   \\
Expt. DFI~\cite{NSG09}   & \hspace{0.25cm} & 86(11)                 & \hspace{0.25cm} &                                             &                                           & \hspace{0.25cm} &                                             &                                             & \hspace{0.25cm} &                                           & \hspace{0.25cm} &                               &                  \\
\mc{14}{c}{Ba$^+$} \\
\vspace{0.1cm}
            & \hspace{0.25cm} & \mc{1}{c}{$6s_{1/2}$}                  & \hspace{0.25cm} & \mc{2}{c}{$5d_{3/2}$}                                                     & \hspace{0.25cm} & \mc{2}{c}{$5d_{5/2}$}                                                        & \hspace{0.25cm} & \mc{1}{c}{$6p_{1/2}$}     & \hspace{0.25cm} & \mc{2}{c}{$6p_{3/2}$} \\
\cline{5-6} \cline{8-9} \cline{13-14}
\vspace{0.1cm}
& \hspace{0.25cm} & \mc{1}{c}{$\alpha_1^{S}$} & \hspace{0.25cm} & \mc{1}{c}{$\alpha_1^{S}$} & \mc{1}{c}{$\alpha_1^{T}$} & \hspace{0.25cm} & \mc{1}{c}{$\alpha_1^{S}$} & \mc{1}{c}{$\alpha_1^{T}$} & \hspace{0.25cm} & \mc{1}{c}{$\alpha_1^{S}$} & \hspace{0.25cm} & \mc{1}{c}{$\alpha_1^{S}$} & \mc{1}{c}{$\alpha_1^{T}$}\\
\hline
DHFCP-LMM                    & \hspace{0.25cm} & 120.74(9)             & \hspace{0.25cm} & 49.438(26)                         & $-$21.403(93)                    & \hspace{0.25cm} & 49.832(28)                            & $-$29.183(120)                  & \hspace{0.25cm} & \;\, 22.39(41)                    & \hspace{0.25cm} & \;\, 45.86(36)       & 3.110(86)   \\
RCCSD~\cite{KSA15}      & \hspace{0.25cm} & 123.18                 & \hspace{0.25cm} & 53.80                                  & $-$22.92                             & \hspace{0.25cm} & 56.53                                     & $-$31.83                            & \hspace{0.25cm} & \;\, 20.46                           & \hspace{0.25cm} & \;\, 45.53              & 4.70           \\
RCCSD(T)~\cite{KSA17} & \hspace{0.25cm} & 123.7(5)               & \hspace{0.25cm} & 54.17(250)                         & $-$22.19(4)                        & \hspace{0.25cm} & 56.87(240)                             & $-$32.17(3)                       & \hspace{0.25cm} &                                           & \hspace{0.25cm} &                              &                   \\
RCCSD(T)~\cite{STD09} & \hspace{0.25cm} & 124.26(100)         & \hspace{0.25cm} & 48.81(46)                           & $-$24.62(28)                      & \hspace{0.25cm} & 50.67(58)                               & $-$30.85(31)                     & \hspace{0.25cm} &                                           & \hspace{0.25cm} &                              &                   \\
RLCCSD(T)~\cite{IS08}   & \hspace{0.25cm} & 124.15                 & \hspace{0.25cm} &                                            &                                            & \hspace{0.25cm} &                                               &                                           & \hspace{0.25cm} &                                           & \hspace{0.25cm} &                              &                   \\
\vspace{0.1cm}
Expt. RESIS~\cite{SL07}  & \hspace{0.25cm} & 123.88(5)            & \hspace{0.25cm} &                                            &                                            & \hspace{0.25cm} &                                               &                                           & \hspace{0.25cm} &                                           & \hspace{0.25cm} &                              &                   \\
\hline
\hline
\end{tabular}
}
\label{table_pola_dipole}
\end{table}

\begin{table}[ht!]
\caption{\small{Static scalar quadrupole $(\alpha_2^{S})$ polarizabilities (in a.u.) of the five lowest states in Ca$^+$, Sr$^+$, and Ba$^+$. Core polarizabilities from~\Ref{JKH83} are added to the DHFCP-LMM valence $\alpha_2^{S}(v)$ results. Comparison with other theory and experiment. Uncertainties in the last digits are given within parentheses.}}
\vspace{0.1cm}
\begin{tabular}{l c l c l c l c l c l}
\hline
\hline
\mc{11}{c}{Ca$^+$} \\
             & \hspace{0.25cm} & \mc{1}{c}{$4s_{1/2}$}         & \hspace{0.25cm} & \mc{1}{c}{$3d_{3/2}$} & \hspace{0.25cm} & \mc{1}{c}{$3d_{5/2}$} & \hspace{0.25cm} & \mc{1}{c}{$4p_{1/2}$} & \hspace{0.25cm} & \mc{1}{c}{$4p_{3/2}$} \\
\vspace{0.1cm}
Method  & \hspace{0.25cm} & \mc{1}{c}{$\alpha_2^{S}$} & \hspace{0.25cm} & \mc{1}{c}{$\alpha_2^{S}$} & \hspace{0.25cm} & \mc{1}{c}{$\alpha_2^{S}$} & \hspace{0.25cm} & \mc{1}{c}{$\alpha_2^{S}$} & \hspace{0.25cm} & \mc{1}{c}{$\alpha_2^{S}$} \\
\hline
DHFCP-LMM                    & \hspace{0.25cm} & 875.78(223) & \hspace{0.25cm} & 5143(111)                   & \hspace{0.25cm} & $-$3435(76)                  & \hspace{0.25cm} & 74803(150)                  & \hspace{0.25cm} & $-$35781(76)                \\
DFCP~\cite{TQS13}        & \hspace{0.25cm} & 882.43         & \hspace{0.25cm} & 4928                            & \hspace{0.25cm} & $-$3304                         & \hspace{0.25cm} & 74660                           & \hspace{0.25cm} & $-$35710                      \\
RCCSD(T)~\cite{ANS12} & \hspace{0.25cm} & 906(5)         & \hspace{0.25cm} &                                     & \hspace{0.25cm} & $-$3706(75)                   & \hspace{0.25cm} &                                      & \hspace{0.25cm} &                                      \\
RMBPT-SD~\cite{SS11}   & \hspace{0.25cm} & 871(4)         & \hspace{0.25cm} &                                     & \hspace{0.25cm} &                                        & \hspace{0.25cm} &                                      & \hspace{0.25cm} &                                       \\
\mc{11}{c}{Sr$^+$} \\
             & \hspace{0.25cm} & \mc{1}{c}{$5s_{1/2}$}         & \hspace{0.25cm} & \mc{1}{c}{$4d_{3/2}$} & \hspace{0.25cm} & \mc{1}{c}{$4d_{5/2}$} & \hspace{0.25cm} & \mc{1}{c}{$5p_{1/2}$} & \hspace{0.25cm} & \mc{1}{c}{$5p_{3/2}$} \\
\vspace{0.1cm}
             & \hspace{0.25cm} & \mc{1}{c}{$\alpha_2^{S}$} & \hspace{0.25cm} & \mc{1}{c}{$\alpha_2^{S}$} & \hspace{0.25cm} & \mc{1}{c}{$\alpha_2^{S}$} & \hspace{0.25cm} & \mc{1}{c}{$\alpha_2^{S}$} & \hspace{0.25cm} & \mc{1}{c}{$\alpha_2^{S}$} \\
\hline
DHFCP-LMM                    & \hspace{0.25cm} & 1351.7(44)   & \hspace{0.25cm} & 2777(43)                     & \hspace{0.25cm} & $-$1773(36)                 & \hspace{0.25cm} & 31576(79)                     & \hspace{0.25cm} & $-$13091(43)                \\
RCICP~\cite{JMC16}       & \hspace{0.25cm} & 1356.3(315) & \hspace{0.25cm} & 2713(44)                     & \hspace{0.25cm} & $-$1728(23)                 & \hspace{0.25cm} & 31596(455)                   & \hspace{0.25cm} & $-$13099(225)              \\
RCCSD(T)~\cite{ANS12} & \hspace{0.25cm} & 1366(9)        & \hspace{0.25cm} &                                    & \hspace{0.25cm} & $-$1732(41)                 & \hspace{0.25cm} &                                       & \hspace{0.25cm} &                                       \\
RMBPT-SD~\cite{Sa10}   & \hspace{0.25cm} & 1370.0(28)   & \hspace{0.25cm} &                                    & \hspace{0.25cm} &                                      & \hspace{0.25cm} &                                       & \hspace{0.25cm} &                                       \\
Expt. DFI~\cite{NSG09}  & \hspace{0.25cm} & 1.1(10)$\times 10^3$ & \hspace{0.25cm} &                      & \hspace{0.25cm} &                                      & \hspace{0.25cm} &                                       & \hspace{0.25cm} &                                       \\
\mc{11}{c}{Ba$^+$} \\
             & \hspace{0.25cm} & \mc{1}{c}{$6s_{1/2}$}         & \hspace{0.25cm} & \mc{1}{c}{$5d_{3/2}$} & \hspace{0.25cm} & \mc{1}{c}{$5d_{5/2}$} & \hspace{0.25cm} & \mc{1}{c}{$6p_{1/2}$} & \hspace{0.25cm} & \mc{1}{c}{$6p_{3/2}$}  \\
\vspace{0.1cm}
             & \hspace{0.25cm} & \mc{1}{c}{$\alpha_2^{S}$} & \hspace{0.25cm} & \mc{1}{c}{$\alpha_2^{S}$} & \hspace{0.25cm} & \mc{1}{c}{$\alpha_2^{S}$} & \hspace{0.25cm} & \mc{1}{c}{$\alpha_2^{S}$} & \hspace{0.25cm} & \mc{1}{c}{$\alpha_2^{S}$} \\
\hline
DHFCP-LMM                   & \hspace{0.25cm} & 4067(4)        & \hspace{0.25cm} & 728.4(15)                     & \hspace{0.25cm} & $-$1127(6)                   & \hspace{0.25cm} & 23423(4)                       & \hspace{0.25cm} & $-$6973(7)                    \\
RLCCSD(T)~\cite{IS08}  & \hspace{0.25cm} & 4182(34)      & \hspace{0.25cm} &                                     & \hspace{0.25cm} &                                      & \hspace{0.25cm} &                                       & \hspace{0.25cm} &                                       \\
\vspace{0.1cm}
Expt. RESIS~\cite{SL07} & \hspace{0.25cm} & 4420(250)   & \hspace{0.25cm} &                                      & \hspace{0.25cm} &                                      & \hspace{0.25cm} &                                       & \hspace{0.25cm} &                                       \\
\hline
\hline
\end{tabular}
\label{table_pola_quadrupole}
\end{table}

\twocolumngrid

\subsubsection{Quadrupole polarizabilities}
\label{subsubsec:res_pola_quadrupole}

The ground $(n+1)s_{1/2}$ quadrupole polarizabilities involve $n'd_{3/2,5/2}$ states. The present values agree very well ($<1\%$ differences) with DFCP~\cite{TQS13} for Ca$^+$ and with RCICP~\cite{JMC16} for Sr$^+$, while no reference value is available with these approaches for Ba$^+$. The comparison with \textit{ab initio} methods shows that the present values are a few percents lower, for the same reason as for dipole polarizabilities. The Sr$^+$ experimental value from \Ref{NSG09} is clearly incompatible with the theoretical works, while for Ba$^+$ the experimental result from \Ref{SL07} is in favor of the \textit{ab initio} number, matching within the experimental uncertainties.

The $nd_j$ quadrupole polarizabilities involve $n's_{1/2},n'd_{3/2,5/2},n'g_{7/2}$ states, and additional $n'g_{9/2}$ states for $nd_{5/2}$. Hence, $nd_j$ intermediate states must be excluded from the sum over $n'$ in the $nd_j$ polarizability. The present values for Ca$^+$ differ from the DFCP result by $4\%$, and do not match with RCICP within the theoretical uncertainties for Sr$^+$. The agreement with other semiempirical approaches is thus poor. \textit{Ab initio} values are only available for $nd_{5/2}$ states. The level of agreement ranges from $5\%$ to $9\%$ for Ca$^+$, and below $5\%$ for Sr$^+$. No reference number is available for Ba$^+$. Hence, around $5\%$ uncertainty should be assigned to the present values for Ba$^+$.

The $(n+1)p_j$ quadrupole polarizabilities involve $n'p_{3/2},n'f_{5/2}$ states, and additional $n'p_{1/2},n'f_{7/2}$ states for $(n+1)p_{3/2}$. Hence, $(n+1)p_{3/2}$ intermediate states must be excluded from the sum over $n'$ in the $(n+1)p_{3/2}$ polarizability. Excellent consistency ($0.2\%$ differences) with DFCP and RCICP is obtained for Ca$^+$, and the values agree very well within the theoretical uncertainties for Sr$^+$. Neither semiempirical nor \textit{ab initio} calculations are available for Ba$^+$.

\subsection{Calculations of decay rates and lifetimes}
\label{subsec:res_2E1_lifetimes}

\begin{table}[ht!]
\caption{\small{$2E1$ decay rates, $W_{2E1}$ (in s$^{-1}$), of the $nd_{j} \rightarrow (n+1)s_{1/2}$ transitions in Ca$^{+}$ ($n=3$), Sr$^{+}$ ($n=4$), and Ba$^{+}$ ($n=5$). DHF-LMM and DHFCP-LMM values are compared with results from Refs.~\cite{SJS10,SJS17}$^{\text{a}}$. Uncertainties in the last digits are given within parentheses. Powers of 10 are indicated within brackets.}}
\vspace{0.1cm}
\resizebox{0.485\textwidth}{!}{			
\begin{tabular}{l c c c l c c c r}
\hline
\hline
\vspace{0.1cm}
                                                   & & \mc{7}{c}{$W_{2E1}$ (s$^{-1}$)} 																			    \\
\cline{3-9}
\vspace{0.1cm}
                                                   & & \mc{3}{c}{LMM}                                  & & \mc{3}{c}{Refs.~\cite{SJS10,SJS17}$^{\text{a}}$} \\
\cline{3-5} \cline{7-9}
\vspace{0.1cm}
Transition                                     & & DHF                 & & \mc{1}{c}{DHFCP} & & DHF                 & & \mc{1}{c}{All order} \\
\hline
\multicolumn{9}{c}{Ca$^{+}$} \\
$3d_{3/2} \rightarrow 4s_{1/2}$ & & 3.446\,[$-$3] & & 1.030(2)\,[$-$4]   & & 3.458\,[$-$3] & & 9.800\,[$-$5]         \\
\vspace{0.2cm}
$3d_{5/2} \rightarrow 4s_{1/2}$ & & 3.392\,[$-$3] & & 1.047(2)\,[$-$4]   & & 3.404\,[$-$3] & & 9.945\,[$-$5]         \\
\multicolumn{9}{c}{Sr$^{+}$} \\
$4d_{3/2} \rightarrow 5s_{1/2}$ & & 2.765\,[$-$3] & & 3.465(11)\,[$-$4] & & 2.777\,[$-$3] & & 3.525\,[$-$4]         \\
\vspace{0.2cm}
$4d_{5/2} \rightarrow 5s_{1/2}$ & & 2.704\,[$-$3] & & 3.753(11)\,[$-$4] & & 2.718\,[$-$3] & & 3.807\,[$-$4]         \\
\multicolumn{9}{c}{Ba$^{+}$} \\
$5d_{3/2} \rightarrow 6s_{1/2}$ & & 7.359\,[$-$6] & & 1.446(9)\,[$-$7]   & & 7.384\,[$-$6] & & 1.538\,[$-$7]         \\
\vspace{0.1cm}
$5d_{5/2} \rightarrow 6s_{1/2}$ & & 1.005\,[$-$5] & & 3.851(25)\,[$-$7] & & 1.013\,[$-$5] & & 4.039\,[$-$7]         \\
\hline
\hline			 				 	 
\end{tabular}
}
\vspace{-0.3cm}
\begin{flushleft}
\small{$^{\text{a}}$Corrected values~\cite{SJS17}; a factor 1/2 is missing in \Ref{SJS10}.}
\end{flushleft}
 \label{table_2E1_two_photon}
\end{table}

\tablename{~\ref{table_2E1_two_photon}} displays $2E1$ decay rates, $W_{2E1}$ (in s$^{-1}$), of the $nd_{j} \rightarrow (n+1)s_{1/2}$ transitions in Ca$^{+}$ ($n=3$), Sr$^{+}$ ($n=4$), and Ba$^{+}$ ($n=5$) ions. The $nd_{3/2} \rightarrow (n+1)s_{1/2}$ transitions involve $n'p_{1/2,3/2}$ intermediate states while the $nd_{5/2} \rightarrow (n+1)s_{1/2}$ transitions only involve $n'p_{3/2}$ states. The DHF-LMM and DHFCP-LMM results are computed in the length gauge with $N_c=N_v=N'=50$ Lagrange functions. The integral over $\omega_1$ is evaluated with $N_{\omega_1}=50$ mesh points. Significant digits are estimated by increasing $N'$ from 50 to 80 and by comparing results from M I and M II. Better agreement between M~I and M~II is obtained for DHFCP-LMM results, since DHF-LMM energies differ more significantly between both integration methods. As for \tablename{s~\ref{table_pola_dipole}} and~\ref{table_pola_quadrupole}, using the second set of core dipole polarizabilities enables to estimate theoretical uncertainties on the DHFCP-LMM values displayed in \tablename{~\ref{table_2E1_two_photon}}.

As mentioned in Sec.~\ref{sec:intro}, only one prior calculation has been carried out in these ions~\cite{SJS10}, using the \textit{ab initio} relativistic single-double all-order method. The comparison with the present results enabled to detect that a factor 1/2 was missing in Table I of \Ref{SJS10}. Indeed, both approaches led to comparable differential decay rates, $d\overline{W}_{2E1}/d\omega_1$, and the only source of error was a wrong choice of bounds for the integration over $\omega_1$. A corrigendum has recently been published in \Ref{SJS17}.

The DHF-LMM results agree very well with corrected reference values, since both calculations are based on an \textit{ab initio} method. The differences range from $0.4\%$ to $0.5\%$, apart from $0.8\%$ for the $5d_{5/2} \rightarrow 6s_{1/2}$ transition in Ba$^+$. By contrast, differences between the DHFCP-LMM results and corrected all-order values are one order of magnitude higher, ranging from $1.4\%$ to $6.0\%$ depending on the studied ion. However, overall agreement is highly satisfying, keeping in mind that semiempirical results are compared to an \textit{ab initio} method that explicitly includes single and double electron excitations to all orders of perturbation theory. Besides, the present study leads to the same conclusion as in \Ref{SJS10}, i.e., that the DHF values of the $2E1$ decay rates are strongly modified by the inclusion of electron correlation. Indeed, the DHF calculation overestimates the rates by factors of $10-50$.

\tablename{~\ref{table_lifetimes}} displays lifetimes, $\tau$ (in s), and multipole contributions to the transition rates, $W$ (in~s$^{-1}$), of the $nd_{j}$ states in Ca$^{+}$ ($n=3$), Sr$^{+}$ ($n=4$), and Ba$^{+}$ ($n=5$) ions. The DHFCP-LMM values of the $2E1$ decay rates are taken from \tablename{~\ref{table_2E1_two_photon}}. The DHFCP-LMM values of the $E2$ decay rates are computed in the length gauge with $N_c=N_v=50$, and $M1$ decay rates are computed with the same parameters. Significant digits of the $E2$ and $M1$ results are estimated by comparison with $N_v=60$ considering M~I and M~II, and the comparison of the results from M~I and M~II allows to assess the precision of the values displayed in \tablename{~\ref{table_lifetimes}}. Again, an estimation of theoretical uncertainties on the values displayed in \tablename{~\ref{table_lifetimes}} is obtained by using the second set of core dipole and quadrupole polarizabilities.

DHFCP-LMM results of the total lifetimes $\tau$ are compared with other theory and with observation in order to study the accuracy of the present approach. While the cited theoretical references only report on calculations of $E2$ and $M1$ contributions to the lifetimes of the $nd_{j}$ states, the present work also includes the $2E1$ contributions.

While the $nd_{3/2}$ states can only decay via $nd_{3/2} \rightarrow (n+1)s_{1/2}$ channels, the $nd_{5/2}$ states decay via $nd_{5/2} \rightarrow (n+1)s_{1/2}$ and $nd_{5/2} \rightarrow nd_{3/2}$ channels. $2E1$ decay rates being proportional to $\omega^6$, their contribution is negligible for $nd_{5/2} \rightarrow nd_{3/2}$ ($<10^{-13}$ s$^{-1}$) but not for $nd_{j} \rightarrow (n+1)s_{1/2}$. $E2$ decay rates, proportional to $\omega^5$, are dominant for $nd_{j} \rightarrow (n+1)s_{1/2}$. They are negligible for $nd_{5/2} \rightarrow nd_{3/2}$ in Ca$^+$ and Sr$^+$, but they become comparable to $2E1$ contributions for $5d_{5/2} \rightarrow 5d_{3/2}$ in Ba$^+$. $M1$ decay rates, proportional to $\omega^3$, are negligible for $nd_{3/2} \rightarrow (n+1)s_{1/2}$ but not for $nd_{5/2} \rightarrow nd_{3/2}$. However, their contribution is only significant to the $5d_{3/2}$ lifetime in Ba$^+$.

Taking the inverse of the total decay rates $W$ (in~s$^{-1}$) yields the total lifetimes $\tau$ (in~s). The DHFCP-LMM results agree very well ($<1\%$ differences) with DFCP~\cite{TQS13} for Ca$^+$ and RCICP~\cite{JMC16} for Sr$^+$, and good consistency is found with various \textit{ab initio} calculations for these two ions. Among them, $0.5\%-5\%$ differences are obtained with the very recent work using the relativistic all-order method~\cite{SSJ17}. For each $nd_j$ lifetime of Ca$^+$ and Sr$^+$, the present results lie within the uncertainties of at least one experiment, and the discrepancies with the other experimental values are not high. It should be noticed that discrepancies also occur among theoretical and experimental values, as well as between the two of them. Lifetimes are longer in Ba$^+$ than in Ca$^+$ and Sr$^+$. Comparison of the present calculation of $5d_j$ lifetimes with \Ref{SSJ17} yields $0.9\%-5\%$ differences, in the same range as in Ca$^+$ and Sr$^+$. Values vary from one to two units between \textit{ab initio} calculations, and the large experimental uncertainties due to the long lifetimes do not allow to discriminate between the different works.


\vspace{-0.25cm}

\section{Conclusions}
\label{sec:conc}

This work presents DHFCP-LMM calculations of polarizabilities, one- and two-photon decay rates, and associated lifetimes in Ca$^{+}$, Sr$^{+}$, and Ba$^{+}$ ions. Two integration methods are devised to compute two-electron Slater integrals, and the comparison of their results allows to assess the precision of the values displayed in the tables.  \pagebreak

\onecolumngrid

\begin{table}[ht!]
\caption{\small{Lifetimes, $\tau$ (in s), and multipole contributions to the transition rates, $W$ (in s$^{-1}$), of the $nd_{j}$ states in Ca$^{+}$ ($n=3$), Sr$^{+}$ ($n=4$), and Ba$^{+}$ ($n=5$). DHFCP-LMM results of the total lifetimes $\tau$ are compared with other theory and experiment. Uncertainties in the last digits are given within parentheses. Powers of 10 are indicated within brackets for $W$.}}
\vspace{0.1cm}
\begin{tabular}{l c c c l c l c l c l c l}
\hline
\hline
\vspace{0.1cm}
                    & \hspace{0.5cm} & \mc{5}{c}{$\tau$ (s)}                                                                                                                                     & \hspace{0.5cm} &\mc{5}{c}{$W$ (s$^{-1}$)}                                                                                                                          \\
\cline{3-7} \cline{9-13}
\vspace{0.1cm}
State            & \hspace{0.5cm} & DHFCP-LMM & \hspace{0.25cm} & \mc{1}{c}{Other theory}    & \hspace{0.25cm} & \mc{1}{c}{Experiment}       & \hspace{0.5cm} & Decay channel                               & \hspace{0.25cm} & Multipole   & \hspace{0.25cm} & \mc{1}{c}{DHFCP-LMM} \\
\hline
 & & \multicolumn{11}{c}{Ca$^{+}$} \\
$3d_{3/2}$ & \hspace{0.5cm} & 1.154(7)     & \hspace{0.25cm} & 1.194(11)~\cite{SSJ17}    & \hspace{0.25cm} & 1.111(46)~\cite{KVV95}   & \hspace{0.5cm} & $3d_{3/2} \rightarrow 4s_{1/2}$ & \hspace{0.25cm} & $E2$         & \hspace{0.25cm} & 8.662(49)\,[$-$1]   \\ 
                   & \hspace{0.5cm} &                    & \hspace{0.25cm} & 1.143(1)~\cite{TQS13}      & \hspace{0.25cm} & 1.17(5)~\cite{LAN99}       & \hspace{0.5cm} &                                                     & \hspace{0.25cm} & $M1$        & \hspace{0.25cm} & 1.947(61)\,[$-$11] \\
                   & \hspace{0.5cm} &                    & \hspace{0.25cm} & 1.185(7)~\cite{SID06}       & \hspace{0.25cm} & 1.20(1)~\cite{BDL00}        & \hspace{0.5cm} &                                                     & \hspace{0.25cm} & $2E1$       & \hspace{0.25cm} & 1.030(2)\,[$-$4]    \\
                   & \hspace{0.5cm} &                    & \hspace{0.25cm} & 1.196(11)~\cite{KBL05}    & \hspace{0.25cm} & 1.176(11)~\cite{KBL05}    & \hspace{0.5cm} &                                                     & \hspace{0.25cm} & $\sum W$ & \hspace{0.25cm} & 8.663(49)\,[$-$1]  \\
\vspace{0.1cm}
                   & \hspace{0.5cm} &                    & \hspace{0.25cm} & 1.16~\cite{VGF92}             & \hspace{0.25cm} & 1.113(45)~\cite{ABG94}   & \hspace{0.5cm} &                                                     & \hspace{0.25cm} &                 & \hspace{0.25cm} &                                \\
$3d_{5/2}$ & \hspace{0.5cm} & 1.124(6)     & \hspace{0.25cm} & 1.163(11)~\cite{SSJ17}     & \hspace{0.25cm} & 1.174(10)~\cite{GHL15}   & \hspace{0.5cm} & $3d_{5/2} \rightarrow 4s_{1/2}$ & \hspace{0.25cm} & $E2$         & \hspace{0.25cm} & 8.892(49)\,[$-$1]   \\
                   & \hspace{0.5cm} &                    & \hspace{0.25cm} & 1.114(1)~\cite{TQS13}      & \hspace{0.25cm} & 1.09(5)~\cite{LAN99}       & \hspace{0.5cm} &                                                     & \hspace{0.25cm} & $2E1$       & \hspace{0.25cm} & 1.047(2)\,[$-$4]     \\
                   & \hspace{0.5cm} &                    & \hspace{0.25cm} & 1.110(9)~\cite{SID06}       & \hspace{0.25cm} & 1.168(7)~\cite{BDL00}      & \hspace{0.5cm} & $3d_{5/2} \rightarrow 3d_{3/2}$ & \hspace{0.25cm} & $E2$         & \hspace{0.25cm} & 2.207(28)\,[$-$13] \\
                   & \hspace{0.5cm} &                    & \hspace{0.25cm} & 1.165(11)~\cite{KBL05}    & \hspace{0.25cm} & 1.168(9)~\cite{KBL05}      & \hspace{0.5cm} &                                                     & \hspace{0.25cm} & $M1$        & \hspace{0.25cm} & 2.422(11)\,[$-$6]    \\
\vspace{0.2cm}
                   & \hspace{0.5cm} &                    & \hspace{0.25cm} & 1.14~\cite{VGF92}            & \hspace{0.25cm} & 1.100(18)~\cite{BRS99}    & \hspace{0.5cm} &                                                      & \hspace{0.25cm} & $\sum W$ & \hspace{0.25cm} & 8.893(49)\,[$-$1]   \\
 & & \multicolumn{11}{c}{Sr$^{+}$} \\
$4d_{3/2}$ & \hspace{0.5cm} & 0.445(3)     & \hspace{0.25cm} & 0.437(14)~\cite{SSJ17}    & \hspace{0.25cm} & 0.435(4)~\cite{MLN99}     & \hspace{0.5cm} & $4d_{3/2} \rightarrow 5s_{1/2}$  & \hspace{0.25cm} & $E2$         & \hspace{0.25cm} & 2.245(18)                \\
                   & \hspace{0.5cm} &                    & \hspace{0.25cm} & 0.4442(67)~\cite{JMC16}  & \hspace{0.25cm} & 0.435(4)~\cite{BMN00}     & \hspace{0.5cm} &                                                     & \hspace{0.25cm} & $M1$         & \hspace{0.25cm} & 9.223(27)\,[$-$11]  \\
                   & \hspace{0.5cm} &                    & \hspace{0.25cm} & 0.441(3)~\cite{JAS09}      & \hspace{0.25cm} & 0.455(29)~\cite{BMN00}   & \hspace{0.5cm} &                                                      & \hspace{0.25cm} & $2E1$       & \hspace{0.25cm} & 3.465(11)\,[$-$4]    \\
\vspace{0.1cm}
                   & \hspace{0.5cm} &                    & \hspace{0.25cm} & 0.426(8)~\cite{SID06}       & \hspace{0.25cm} & 0.395(38)~\cite{GHW87}  & \hspace{0.5cm} &                                                      & \hspace{0.25cm} & $\sum W$ & \hspace{0.25cm} & 2.245(19)                \\
$4d_{5/2}$ & \hspace{0.5cm} & 0.398(3)     & \hspace{0.25cm} & 0.3945(22)~\cite{SSJ17}  & \hspace{0.25cm} & 0.372(25)~\cite{MS90}      & \hspace{0.5cm} & $4d_{5/2} \rightarrow 5s_{1/2}$  & \hspace{0.25cm} & $E2$         & \hspace{0.25cm} & 2.509(20)               \\
                   & \hspace{0.5cm} &                    & \hspace{0.25cm} & 0.3974(59)~\cite{JMC16}  & \hspace{0.25cm} & 0.408(22)~\cite{BMN00}   & \hspace{0.5cm} &                                                      & \hspace{0.25cm} & $2E1$       & \hspace{0.25cm} & 3.753(11)\,[$-$4]   \\
                   & \hspace{0.5cm} &                    & \hspace{0.25cm} & 0.394(3)~\cite{JAS09}      & \hspace{0.25cm} & 0.3908(16)~\cite{LWG05} & \hspace{0.5cm} & $4d_{5/2} \rightarrow 4d_{3/2}$  & \hspace{0.25cm} & $E2$         & \hspace{0.25cm} & 1.130(19)\,[$-$9]   \\
                   & \hspace{0.5cm} &                    & \hspace{0.25cm} & 0.357(12)~\cite{SID06}     & \hspace{0.25cm} & 0.347(11)~\cite{BEG93}    & \hspace{0.5cm} &                                                      & \hspace{0.25cm} & $M1$        & \hspace{0.25cm} & 2.378(1)\,[$-$4]     \\
\vspace{0.2cm}
                   & \hspace{0.5cm} &                    & \hspace{0.25cm} &                                          & \hspace{0.25cm} & 0.345(33)~\cite{GHW87}   & \hspace{0.5cm} &                                                      & \hspace{0.25cm} & $\sum W$ & \hspace{0.25cm} & 2.510(20)               \\
 & & \multicolumn{11}{c}{Ba$^{+}$} \\
$5d_{3/2}$ & \hspace{0.5cm} & 83.86(15)   & \hspace{0.25cm} & 81.4(14)~\cite{SSJ17}       & \hspace{0.25cm} & 79.8(46)~\cite{YND97}      & \hspace{0.5cm} & $5d_{3/2} \rightarrow 6s_{1/2}$  & \hspace{0.25cm} & $E2$         & \hspace{0.25cm} & 1.192(3)\,[$-$2]     \\
                   & \hspace{0.5cm} &                    & \hspace{0.25cm} & 81.5(12)~\cite{IS08}         & \hspace{0.25cm} & 89.4(156)~\cite{GBB07}     & \hspace{0.5cm} &                                                      & \hspace{0.25cm} & $M1$        & \hspace{0.25cm} & 2.696(26)\,[$-$11] \\
                   & \hspace{0.5cm} &                    & \hspace{0.25cm} & 82.0~\cite{GBB07}            & \hspace{0.25cm} &                                           & \hspace{0.5cm} &                                                      & \hspace{0.25cm} & $2E1$       & \hspace{0.25cm} & 1.446(9)\,[$-$7]    \\
                   & \hspace{0.5cm} &                    & \hspace{0.25cm} & 80.086(714)~\cite{SID06} & \hspace{0.25cm} &                                           & \hspace{0.5cm} &                                                     & \hspace{0.25cm} & $\sum W$ & \hspace{0.25cm} & 1.192(3)\,[$-$2]     \\
\vspace{0.1cm}
                   & \hspace{0.5cm} &                    & \hspace{0.25cm} & 81.5~\cite{DFG01}             & \hspace{0.25cm} &                                          & \hspace{0.5cm} &                                                      & \hspace{0.25cm} &                 & \hspace{0.25cm} &                                \\
$5d_{5/2}$ & \hspace{0.5cm} & 31.09(4)     & \hspace{0.25cm} & 30.34(48)~\cite{SSJ17}     & \hspace{0.25cm} & 31.2(9)~\cite{ANH14}       & \hspace{0.5cm} & $5d_{5/2} \rightarrow 6s_{1/2}$  & \hspace{0.25cm} & $E2$         & \hspace{0.25cm} & 2.662(5)\,[$-$2]     \\
                   & \hspace{0.5cm} &                    & \hspace{0.25cm} & 30.3(4)~\cite{IS08}            & \hspace{0.25cm} & 34.5(35)~\cite{MS90}       & \hspace{0.5cm} &                                                      & \hspace{0.25cm} & $2E1$       & \hspace{0.25cm} & 3.851(25)\,[$-$7]   \\
                   & \hspace{0.5cm} &                    & \hspace{0.25cm} & 31.6~\cite{GBB07}             & \hspace{0.25cm} & 32.0(46)~\cite{GBB07}      & \hspace{0.5cm} & $5d_{5/2} \rightarrow 5d_{3/2}$  & \hspace{0.25cm} & $E2$         & \hspace{0.25cm} & 2.622(10)\,[$-$7]   \\
                   & \hspace{0.5cm} &                    & \hspace{0.25cm} & 29.856(296)~\cite{SID06} & \hspace{0.25cm} & 32(5)~\cite{NSD86}           & \hspace{0.5cm} &                                                      & \hspace{0.25cm} & $M1$        & \hspace{0.25cm} & 5.543(2)\,[$-$3]     \\
\vspace{0.1cm}
                   & \hspace{0.5cm} &                    & \hspace{0.25cm} & 30.3~\cite{DFG01}             & \hspace{0.25cm} &                                          & \hspace{0.5cm} &                                                      & \hspace{0.25cm} & $\sum W$ & \hspace{0.25cm} & 3.216(5)\,[$-$2]     \\
\hline
\hline
\end{tabular}
 \label{table_lifetimes}
\end{table}

\twocolumngrid

\noindent In addition, the effect of variations in the values of the core dipole and quadrupole polarizabilities on the final results is studied, which enables to estimate theoretical uncertainties on the latter.

The core orbitals are defined by a closed-shell DHF calculation with the \textsc{grasp{\small 2}k} package, and are projected on Lagrange bases. The single valence electron is described in the frozen-core approximation by a Dirac-like Hamiltonian involving a CP potential to simulate the core-valence electron correlation. Comparing with \textsc{grasp{\small 2}k} results, the accuracy on core energies is $\sim 10^{-7}$, while the one on DHF valence energies is $\sim 10^{-5}$ for the five lowest states of each ion. With the inclusion of $V_{\text{CP}}$, calculated energies are fitted with relative errors $<10^{-7}$ in comparison with observation.

Turning to dipole and quadrupole polarizabilities, the agreement with other semiempirical approaches is excellent for Ca$^+$ and Sr$^+$, while no such reference value exists for Ba$^+$. Overall good agreement is obtained with \textit{ab initio} methods and observation, although semiempirical approaches underestimate the ground-state polarizabilities by a few percents. The principal limitation of the accuracy on polarizabilities with such approaches lies in the accuracy of the core polarizabilities, that has to be computed with an independent method.

For the $2E1$ $nd_{j} \rightarrow (n+1)s_{1/2}$ decay rates, a comparison with Refs.~\cite{SJS10,SJS17} shows that both DHF results agree very well with each other, and that a satisfying agreement is obtained between the DHFCP-LMM and all-order values. Both works conclude that the DHF values are strongly modified by the inclusion of electron correlation.

The lifetimes estimation of the metastable $nd_j$ states involves the study of the competition between the $E2$, $M1$, and $2E1$ decay channels. The present results agree very well with other semiempirical approaches for Ca$^+$ and Sr$^+$, and overall good consistency is found with \textit{ab initio} calculations and experiments. Results vary more significantly between calculations for Ba$^+$, and the few existing experiments are associated with large uncertainties. Moreover, the contribution of the $2E1$ processes to the total decay rates of $nd_j$ states is negligible ($0.001-0.01\%$) at the present level of theoretical and experimental accuracy. This conclusion had already been done in \Ref{SJS10}. Estimating more accurate lifetimes for the $nd_j$ metastable states in these three ions currently represents a difficult task. Other decay processes are expected to compete with the $E2$, $M1$, and $2E1$ channels, such as magnetic-field induced transitions (MIT), and hyperfine induced transitions (HFI) for odd-$A$ isotopes of these three ions. These processes are likely to modify the existing theoretical lifetimes values. Besides, new high-precision experimental results are urgently needed to test the theoretical predictions of the $5d_j$ lifetimes in Ba$^+$.

Our work is based on a fully relativistic version of the semiempirical-core-potential approach. As such, it is an approximate method, where the comparison with experiments and other theories should in principle provide an assessment of the errors due to physical effects that are not included in the model. As illustrated by \tablename{s~\ref{table_pola_dipole}}-\ref{table_lifetimes}, estimating more realistic theoretical uncertainties based on such comparisons would be statistically meaningless, due to the dispersion of \textit{ab initio} and experimental values.

Comparison with other theory and observation shows that the DHFCP-LMM method provides a simple and efficient way for evaluating properties of alkali-like ions involving an infinite number of intermediates states, such as relativistic polarizabilities and two-photon decay rates. For the first time, a semiempirical-core-potential calculation of two-photon decay rates is performed, and results from relativistic computations in the Ba$^+$ ion are reported with such an approach. By using the LMM, which allows a simple computation of one-body matrix elements, and by developing Gauss-quadrature-based methods to accurately evaluate the two-electron Slater integrals, precise results are obtained with small computing times and memory requirements. Besides, the use of the \textsc{grasp{\small 2}k} package for core orbitals calculations reduces the code-development effort to only single valence-electron calculations. The present approach can play a role in further improvement of theoretical $nd_j$ lifetimes in these three ions. It can also be used to study a variety of heavy alkali-like systems, such as Cs, Fr, Ra$^+$ and Yb$^+$, for which theoretical results and experimental data are available for comparison, or others for which information is not available. Dynamic polarizabilities, hyperpolarizabilities and dispersion coefficients involved in long-range interactions between pairs of atoms, can be studied in various alkali-like systems. From a methodological point of view, the LMM could also offer some computational advantages for estimating other properties involving an infinite number of intermediate states, such as atomic electric dipole moments and parity nonconservation amplitudes.


\begin{acknowledgments}
This work has been supported by the Belgian F.R.S.-FNRS Fonds de la Recherche Scientifique (CDR J.0047.16), and the BriX IAP Research Program No. P7/12. L.F. acknowledges the support from the FRIA.
\end{acknowledgments}


\appendix*

\section{Two-electron Slater integrals on Lagrange meshes}

Starting with M~I, let us introduce in \Eq{eq_diff_Yk} the expansion $Y_q^k(jc; r) = \bar{h}^{-1/2} \sum_{l=1}^{\bar{N}} y_l \, \hat{f}_l^{(\bar{\alpha}=0)}(r/\bar{h})$, satisfying the boundary condition $Y_q^k(jc; 0)=0$ since $\hat{f}_l^{(0)}(0)=0$. Projecting the l.h.s of \Eq{eq_diff_Yk} on $\bar{h}^{-1/2}\hat{f}_{l'}^{(0)}(r/\bar{h})$ leads to~\cite{Ba15}
\beq
& & \bar{h}^{-2} \left\{ \sum_{l \neq l'}^{\bar{N}} y_{l} \left[ (-1)^{l-l'+1} \, \frac{\bar{x}_l+\bar{x}_{l'}}{\sqrt{\bar{x}_l \bar{x}_{l'}} (\bar{x}_l-\bar{x}_{l'})^2} \right] \right. \eol
& & \left. + y_{l'} \left[ \frac{\bar{x}_{l'}^2 - 2(2\bar{N}+1) \bar{x}_{l'} - 4}{12 \bar{x}_{l'}^2} - \frac{k(k+1)}{\bar{x}_{l'}^2} \right] \right\}
\eeqn{eq_Yk_left_Lag}
for $l'=1$ to $\bar{N}$, using a Gauss quadrature with parameters $\bar{N}=N_v+N_c$, $\bar{h}=2h_c$ and $\bar{\alpha}=0$. The first two values are deduced from expression~\rref{eq_Yk_vc}, while the Schr\"odinger-like form of \Eq{eq_diff_Yk} requires an integer value of $\bar{\alpha}$ to reproduce the exact behavior of $Y_q^k(jc; r)$ near the origin. Projecting the r.h.s. of \Eq{eq_diff_Yk} on $\bar{h}^{-1/2}\hat{f}_{l'}^{(0)}(r/\bar{h})$ leads to
\beq
& & - (\bar{h} h_v)^{-1/2} \sum_{m=1}^{N_G} \tilde{\lambda}_m \, \hat{f}_{l'}^{(0)}(\tilde{h} \tilde{x}_m/\bar{h}) \eol
& & \times \frac{2k+1}{\tilde{x}_m} \, Q_{c}(\tilde{h} \tilde{x}_m) \hat{f}_i^{(\alpha_v)}(\tilde{h} \tilde{x}_m/h_v)
\eeqn{eq_Yk_right_Lag}
for $l'=1$ to $\bar{N}$, using a Gauss quadrature with parameters $N_G > N_v+N_c$, $\tilde{h}=4h_vh_c/(3h_v+2h_c)$ and $\tilde{\alpha}=(\alpha_v+\alpha_c)/2$. The equality of Eqs.~\rref{eq_Yk_left_Lag} and \rref{eq_Yk_right_Lag} defines an $\bar{N} \times \bar{N}$ algebraic system which is solved with a standard technique. For the case $k=0$, the expansion of $Y_q^0(jc; r)$ is not able to reproduce this asymptotic behavior $Y_q^0(jc;\infty)\neq 0$ since $\hat{f}_l^{(0)}(r/\bar{h}) \rightarrow 0$ as $r \rightarrow \infty$. To overcome this issue, the function $Y_q^0(jc;\infty) \, (1-e^{-r})$ is subtracted from $Y_q^0(jc; r)$ in \Eq{eq_diff_Yk}, and \Eq{eq_Yk_right_Lag} is modified to include the Gauss quadrature of $Y_q^0(jc;\infty) \int_0^\infty \bar{h}^{-1/2} \hat{f}_{l'}^{(0)}(r/\bar{h}) \, e^{-r} \, dr$. $Y_q^0(jc; r)$ is recovered by adding back $Y_q^0(jc;\infty) \, (1-e^{-r})$ to the solution of the modified algebraic system. Once $Y_q^k(jc; r)$ is known, integral~\rref{eq_Rk_Vexc_2} is expressed using a Gauss quadrature with the same parameters as in~\rref{eq_Yk_right_Lag}:
\beq
R^k_{pq}(ic,cj) & \approx & h_v^{-1/2} \sum_{m=1}^{N_G} \tilde{\lambda}_m \, \hat{f}_i^{(\alpha_v)}(\tilde{h} \tilde{x}_m/h_v) P_{c}(\tilde{h} \tilde{x}_m) \eol
& & \times \frac{1}{\tilde{x}_m} \, Y_q^k(jc; \tilde{h} \tilde{x}_m),
\eeqn{eq_Rk_Vexc_Lag}

Turning to M~II, let us start with the integral over $r$ in \Eq{eq_integral_4}. Defining $\bar{h} \equiv 2h_{v}h_{c}/(h_{v}+h_{c})$, the integrand reads
\beq
\hspace{-0.25cm} (r+r_{2})^{\frac{\alpha_{v}+\alpha_{c}}{2} - k}\, \mathcal{P}_{N_v+N_c-1}(r+r_{2}) \, e^{-(r+r_{2})/\bar{h}},
\eeqn{eq:Integr}	
where $\mathcal{P}_{N_v+N_c-1}$ is a polynomial of degree $N_v+N_c-1$ and the first factor is non-polynomial. Expression \eqref{eq:Integr} suggests that computing the integral over $r$ by the Gauss-Laguerre quadrature with weight function $\exp{(-r/\bar{h})}$ and $N_G>(N_v+N_c)/2$ points should be efficient. Denoting the abscissae and weights of this Gauss quadrature by $\bar{h} \bar{x}_m$ and $\bar{h} \bar{\lambda}_m$ (with $m=1$ to $N_G$), one has
\beq
& & \int_{0}^{\infty} \frac{\hat{f}^{(\alpha_{v})}_{i}[(r+r_{2})/h_{v}] P_c(r+r_{2})}{(r+r_{2})^{k+1}} \, dr \eol
& & \hspace{-0.5cm} \approx \bar{h} \sum_{m=1}^{N_{G}} \bar{\lambda}_{m} \frac{\hat{f}^{(\alpha_{v})}_{i}[(\bar{h}\bar{x}_{m}+r_{2})/h_{v}] P_c(\bar{h}\bar{x}_{m}+r_{2})}{(\bar{h}\bar{x}_{m}+r_{2})^{k+1}}.
\eeqn{eq_int_Gauss_r}
For each term of the sum over index $m$, the integrand of the integral over $r_{2}$ can be written as
\beq
(\bar{h}\bar{x}_{m}+r_{2})^{\frac{\alpha_{v}+\alpha_{c}}{2} - k} \, r _{2}^{\frac{\alpha_{v}+\alpha_{c}}{2}} \, \tilde{\mathcal{P}}_{2 N_v+2 N_c+k-1}(r_2) \, e^{-r_{2}/\tilde{h}}, \eol
\eeqn{eq_int_Gauss_r2}
where $\tilde{\mathcal{P}}_{2 N_v+2 N_c+k-1}$ is a polynomial of degree $2 N_v+2 N_c+k-1$ and $\tilde{h}\equiv\bar{h}/2$. This expression incites us to evaluate the integral over $r_2$ by a Gauss-Laguerre quadrature with weight function $r_{2}^{(\alpha_{v}+\alpha_{c})/2}  \exp{(- r/\tilde{h})}$ and $N'_G>N_v+N_c+k/2$ points. Let us denote by $\tilde{h} \tilde{x}_{m'}$ and $\tilde{h} \tilde{\lambda}_{m'}$ (with $m'=1$ to $N'_G$), the abscissae and weights of this Gauss quadrature. The integral $\mathcal{R}_{pq}^k(ic,cj)$ given by \Eq{eq_integral_4} reads, finally, 
\beq
& & \mathcal{R}_{pq}^k(ic,cj) \approx h_v^{-1} \bar{h} \tilde{h} \sum^{N'_{G}}_{m'=1} \tilde{\lambda}_{m'} \hat{f}^{(\alpha_{v})}_{j}(\tilde{h} \tilde{x}_{m'}/h_{v}) Q_c(\tilde{h} \tilde{x}_{m'}) \eol
& & \times (\tilde{h} \tilde{x}_{m'})^{k} \sum_{m=1}^{N_{G}} \bar{\lambda}_{m} \frac{\hat{f}^{(\alpha_{v})}_{i}[(\bar{h} \bar{x}_{m}+\tilde{h} \tilde{x}_{m'})/h_{v}] P_c(\bar{h} \bar{x}_{m}+\tilde{h} \tilde{x}_{m'})}{(\bar{h} \bar{x}_{m}+\tilde{h} \tilde{x}_{m'})^{k+1}}. \eol
\eeqn{eq_R1_Gauss}


\twocolumngrid

\onecolumngrid


\begin{thebibliography}{76}%
\makeatletter
\providecommand \@ifxundefined [1]{%
 \@ifx{#1\undefined}
}%
\providecommand \@ifnum [1]{%
 \ifnum #1\expandafter \@firstoftwo
 \else \expandafter \@secondoftwo
 \fi
}%
\providecommand \@ifx [1]{%
 \ifx #1\expandafter \@firstoftwo
 \else \expandafter \@secondoftwo
 \fi
}%
\providecommand \natexlab [1]{#1}%
\providecommand \enquote  [1]{``#1''}%
\providecommand \bibnamefont  [1]{#1}%
\providecommand \bibfnamefont [1]{#1}%
\providecommand \citenamefont [1]{#1}%
\providecommand \href@noop [0]{\@secondoftwo}%
\providecommand \href [0]{\begingroup \@sanitize@url \@href}%
\providecommand \@href[1]{\@@startlink{#1}\@@href}%
\providecommand \@@href[1]{\endgroup#1\@@endlink}%
\providecommand \@sanitize@url [0]{\catcode `\\12\catcode `\$12\catcode `\&12\catcode `\#12\catcode `\^12\catcode `\_12\catcode `\%12\relax}%
\providecommand \@@startlink[1]{}%
\providecommand \@@endlink[0]{}%
\providecommand \url  [0]{\begingroup\@sanitize@url \@url }%
\providecommand \@url [1]{\endgroup\@href {#1}{\urlprefix }}%
\providecommand \urlprefix  [0]{URL }%
\providecommand \Eprint [0]{\href }%
\providecommand \doibase [0]{http://dx.doi.org/}%
\providecommand \selectlanguage [0]{\@gobble}%
\providecommand \bibinfo  [0]{\@secondoftwo}%
\providecommand \bibfield  [0]{\@secondoftwo}%
\providecommand \translation [1]{[#1]}%
\providecommand \BibitemOpen [0]{}%
\providecommand \bibitemStop [0]{}%
\providecommand \bibitemNoStop [0]{.\EOS\space}%
\providecommand \EOS [0]{\spacefactor3000\relax}%
\providecommand \BibitemShut [1]{\csname bibitem#1\endcsname}%
\let\auto@bib@innerbib\@empty
\bibitem [{\citenamefont {Mitroy}\ \emph {et~al.}(2010)\citenamefont {Mitroy}, \citenamefont {Safronova},\ and\ \citenamefont {Clark}}]{MSC10}%
\BibitemOpen
\bibfield  {author} {\bibinfo {author} {\bibfnamefont {J.}~\bibnamefont {Mitroy}}, \bibinfo {author} {\bibfnamefont {M.~S.}\ \bibnamefont {Safronova}}, \ and\ \bibinfo {author} {\bibfnamefont {C.~W.}\ \bibnamefont {Clark}},\ }\href@noop {}
{\bibfield {journal} {\bibinfo {journal} {J. Phys. B}\ }\textbf {\bibinfo {volume} {43}},\ \bibinfo {pages} {202001} (\bibinfo {year} {2010})}\BibitemShut {NoStop}%
\bibitem [{\citenamefont {Ludlow}\ \emph {et~al.}(2015)\citenamefont {Ludlow}, \citenamefont {Boyd}, \citenamefont {Ye}, \citenamefont {Peik},\ and\ \citenamefont {Schmidt}}]{LBY15}%
\BibitemOpen \bibfield  {author} {\bibinfo {author} {\bibfnamefont {A.~D.}\ \bibnamefont {Ludlow}}, \bibinfo {author} {\bibfnamefont {M.~M.}\ \bibnamefont {Boyd}}, \bibinfo {author} {\bibfnamefont {J.}~\bibnamefont {Ye}}, \bibinfo {author} {\bibfnamefont {E.}~\bibnamefont {Peik}}, \ and\ \bibinfo {author} {\bibfnamefont {P.~O.}\ \bibnamefont {Schmidt}},\ }\href@noop {}
{\bibfield {journal} {\bibinfo  {journal} {Rev. Mod. Phys.}\ }\textbf {\bibinfo {volume} {87}},\ \bibinfo {pages} {637} (\bibinfo {year} {2015})}\BibitemShut {NoStop}%
\bibitem [{\citenamefont {Jiang}\ \emph {et~al.}(2017)\citenamefont {Jiang}, \citenamefont {Jiang}, \citenamefont {Wang}, \citenamefont {Zhang}, \citenamefont {Xie},\ and\ \citenamefont {Dong}}]{JJW17}%
\BibitemOpen
\bibfield {author} {\bibinfo {author} {\bibfnamefont {J.}~\bibnamefont {Jiang}}, \bibinfo {author} {\bibfnamefont {L.}~\bibnamefont {Jiang}}, \bibinfo {author} {\bibfnamefont {X.}~\bibnamefont {Wang}}, \bibinfo {author} {\bibfnamefont {D.~H.}\ \bibnamefont {Zhang}}, \bibinfo {author} {\bibfnamefont {L.~Y.}\ \bibnamefont {Xie}},\ and\ \bibinfo {author} {\bibfnamefont {C.~Z.}\ \bibnamefont {Dong}}, }\href@noop {}
{\bibfield {journal} {\bibinfo {journal} {arXiv:1703.09950v1}} (\bibinfo {year} {2017})}\BibitemShut {NoStop}%
\bibitem [{\citenamefont {Tang}\ \emph {et~al.}(2013)\citenamefont {Tang}, \citenamefont {Qiao}, \citenamefont {Shi},\ and\ \citenamefont {Mitroy}}]{TQS13}%
\BibitemOpen
\bibfield {author} {\bibinfo {author} {\bibfnamefont {Y.~B.}\ \bibnamefont {Tang}}, \bibinfo {author} {\bibfnamefont {H.~X.}\ \bibnamefont {Qiao}}, \bibinfo {author} {\bibfnamefont {T.~Y.}\ \bibnamefont {Shi}},\ and\ \bibinfo {author} {\bibfnamefont {J.}~\bibnamefont {Mitroy}},\ }\href@noop {}
{\bibfield {journal} {\bibinfo {journal} {Phys. Rev. A}\ }\textbf {\bibinfo {volume} {87}},\ \bibinfo {pages} {042517} (\bibinfo {year} {2013})}\BibitemShut {NoStop}%
\bibitem [{\citenamefont {Safronova}\ \emph {et~al.}(2011)\citenamefont {Safronova}\ and\ \citenamefont {Safronova}}]{SS11}%
\BibitemOpen
\bibfield {author} {\bibinfo {author} {\bibfnamefont {M.~S.}\ \bibnamefont {Safronova}}\ and\ \bibinfo {author} {\bibfnamefont {U.~I.}\ \bibnamefont {Safronova}},\ }\href@noop {}
{\bibfield {journal} {\bibinfo {journal} {Phys. Rev. A}\ }\textbf {\bibinfo {volume} {83}},\ \bibinfo {pages} {012503} (\bibinfo {year} {2011})}\BibitemShut {NoStop}%
\bibitem [{\citenamefont {Kaur}\ \emph {et~al.}(2015)\citenamefont {Kaur}, \citenamefont {Singh}, \citenamefont {Arora},\ and\ \citenamefont {Sahoo}}]{KSA15}%
\BibitemOpen
\bibfield {author} {\bibinfo {author} {\bibfnamefont {J.}~\bibnamefont {Kaur}}, \bibinfo {author} {\bibfnamefont {S.}~\bibnamefont {Singh}}, \bibinfo {author} {\bibfnamefont {B.}~\bibnamefont {Arora}},\ and\ \bibinfo {author} {\bibfnamefont {B.~K.}\ \bibnamefont {Sahoo}},\ }\href@noop {}
{\bibfield {journal} {\bibinfo {journal} {Phys. Rev. A}\ }\textbf {\bibinfo {volume} {92}},\ \bibinfo {pages} {031402(R)} (\bibinfo {year} {2015})}\BibitemShut {NoStop}%
\bibitem [{\citenamefont {Kaur}\ \emph {et~al.}(2017)\citenamefont {Kaur}, \citenamefont {Singh}, \citenamefont {Arora},\ and\ \citenamefont {Sahoo}}]{KSA17}%
\BibitemOpen
\bibfield {author} {\bibinfo {author} {\bibfnamefont {J.}~\bibnamefont {Kaur}}, \bibinfo {author} {\bibfnamefont {S.}~\bibnamefont {Singh}}, \bibinfo {author} {\bibfnamefont {B.}~\bibnamefont {Arora}},\ and\ \bibinfo {author} {\bibfnamefont {B.~K.}\ \bibnamefont {Sahoo}},\ }\href@noop {}
{\bibfield {journal} {\bibinfo {journal} {Phys. Rev. A}\ }\textbf {\bibinfo {volume} {95}},\ \bibinfo {pages} {042501} (\bibinfo {year} {2017})}\BibitemShut {NoStop}%
\bibitem [{\citenamefont {Arora}\ \emph {et~al.}(2007)\citenamefont {Arora}, \citenamefont {Safronova},\ and\ \citenamefont {Clark}}]{ASC07}%
\BibitemOpen
\bibfield {author} {\bibinfo {author} {\bibfnamefont {B.}~\bibnamefont {Arora}}, \bibinfo {author} {\bibfnamefont {M.~S.}\ \bibnamefont {Safronova}},\ and\ \bibinfo {author} {\bibfnamefont {C.~W.}\ \bibnamefont {Clark}},\ }\href@noop {}
{\bibfield {journal} {\bibinfo {journal} {Phys. Rev. A}\ }\textbf {\bibinfo {volume} {76}},\ \bibinfo {pages} {064501} (\bibinfo {year} {2007})}\BibitemShut {NoStop}%
\bibitem [{\citenamefont {Jiang}\ \emph {et~al.}(2016)\citenamefont {Jiang}, \citenamefont {Mitroy}, \citenamefont {Cheng},\ and\ \citenamefont {Bromley}}]{JMC16}%
\BibitemOpen
\bibfield {author} {\bibinfo {author} {\bibfnamefont {J.}~\bibnamefont {Jiang}}, \bibinfo {author} {\bibfnamefont {J.}~\bibnamefont {Mitroy}}, \bibinfo {author} {\bibfnamefont {Y.}~\bibnamefont {Cheng}},\ and\ \bibinfo {author} {\bibfnamefont {M.~W.~J.}\ \bibnamefont {Bromley}},\ }\href@noop {}
{\bibfield {journal} {\bibinfo {journal} {Phys. Rev. A}\ }\textbf {\bibinfo {volume} {94}},\ \bibinfo {pages} {062514} (\bibinfo {year} {2016})}\BibitemShut {NoStop}%
\bibitem [{\citenamefont {Jiang}\ \emph {et~al.}(2009)\citenamefont {Jiang}, \citenamefont {Arora}, \citenamefont {Safronova},\ and\ \citenamefont {Clark}}]{JAS09}%
\BibitemOpen
\bibfield {author} {\bibinfo {author} {\bibfnamefont {J.}~\bibnamefont {Jiang}}, \bibinfo {author} {\bibfnamefont {B.}~\bibnamefont {Arora}}, \bibinfo {author} {\bibfnamefont {M.~S.}\ \bibnamefont {Safronova}},\ and\ \bibinfo {author} {\bibfnamefont {C.~W.}\ \bibnamefont {Clark}},\ }\href@noop {}
{\bibfield {journal} {\bibinfo {journal} {J. Phys. B: At. Mol. Opt. Phys.}\ }\textbf {\bibinfo {volume} {42}},\ \bibinfo {pages} {154020} (\bibinfo {year} {2009})}\BibitemShut {NoStop}%
\bibitem [{\citenamefont {Iskrenova-Tchoukova}\ \emph {et~al.}(2008)\citenamefont {Iskrenova-Tchoukova}\ and\ \citenamefont {Safronova}}]{IS08}%
\BibitemOpen
\bibfield {author} {\bibinfo {author} {\bibfnamefont {E.}~\bibnamefont {Iskrenova-Tchoukova}}\ and\ \bibinfo {author} {\bibfnamefont {M.~S.}\ \bibnamefont {Safronova}},\ }\href@noop {}
{\bibfield {journal} {\bibinfo {journal} {Phys. Rev. A}\ }\textbf {\bibinfo {volume} {78}},\ \bibinfo {pages} {012508} (\bibinfo {year} {2008})}\BibitemShut {NoStop}%
\bibitem [{\citenamefont {Safronova}(2010)}]{Sa10}%
\BibitemOpen
\bibfield {author} {\bibinfo {author} {\bibfnamefont {U.~I.}\ \bibnamefont {Safronova}},\ }\href@noop {}
{\bibfield {journal} {\bibinfo {journal} {Phys. Rev. A}\ }\textbf {\bibinfo {volume} {82}},\ \bibinfo {pages} {022504} (\bibinfo {year} {2010})}\BibitemShut {NoStop}%
\bibitem [{\citenamefont {Sahoo}\ \emph {et~al.}(2009)\citenamefont {Sahoo}, \citenamefont {Timmermans}, \citenamefont {Das},\ and\ \citenamefont {Mukherjee}}]{STD09}%
\BibitemOpen
\bibfield {author} {\bibinfo {author} {\bibfnamefont {B.~K.}\ \bibnamefont {Sahoo}}, \bibinfo {author} {\bibfnamefont {R.~G.~E.}\ \bibnamefont {Timmermans}}, \bibinfo {author} {\bibfnamefont {B.~P.}\ \bibnamefont {Das}},\ and\ \bibinfo {author} {\bibfnamefont {D.}~\bibnamefont {Mukherjee}},\ }\href@noop {}
{\bibfield {journal} {\bibinfo {journal} {Phys. Rev. A}\ }\textbf {\bibinfo {volume} {80}},\ \bibinfo {pages} {062506} (\bibinfo {year} {2009})}\BibitemShut {NoStop}%
\bibitem [{\citenamefont {Arora}\ \emph {et~al.}(2012)\citenamefont {Arora}, \citenamefont {Nandy},\ and\ \citenamefont {Sahoo}}]{ANS12}%
\BibitemOpen
\bibfield {author} {\bibinfo {author} {\bibfnamefont {B.}~\bibnamefont {Arora}}, \bibinfo {author} {\bibfnamefont {D.~K.}\ \bibnamefont {Nandy}},\ and\ \bibinfo {author} {\bibfnamefont {B.~K.}\ \bibnamefont {Sahoo}},\ }\href@noop {}
{\bibfield {journal} {\bibinfo {journal} {Phys. Rev. A}\ }\textbf {\bibinfo {volume} {85}},\ \bibinfo {pages} {012506} (\bibinfo {year} {2012})}\BibitemShut {NoStop}%
\bibitem [{\citenamefont {Chou}\ \emph {et~al.}(2010)\citenamefont {Chou}, \citenamefont {Hume}, \citenamefont {Koelemeij}, \citenamefont {Wineland},\ and\ \citenamefont {Rosenband}}]{CHK10}%
\BibitemOpen \bibfield  {author} {\bibinfo {author} {\bibfnamefont {C.~W.}\ \bibnamefont {Chou}}, \bibinfo {author} {\bibfnamefont {D.~B.}\ \bibnamefont {Hume}}, \bibinfo {author} {\bibfnamefont {J.~C.~J.}\ \bibnamefont {Koelemeij}}, \bibinfo {author} {\bibfnamefont {D.~J.}\ \bibnamefont {Wineland}}, \ and\ \bibinfo {author} {\bibfnamefont {T.}~\bibnamefont {Rosenband}},\ }\href@noop {}
{\bibfield {journal} {\bibinfo  {journal} {Phys. Rev. Lett.}\ }\textbf {\bibinfo {volume} {104}},\ \bibinfo {pages} {070802} (\bibinfo {year} {2010})}\BibitemShut {NoStop}%
\bibitem [{\citenamefont {Huntemann}\ \emph {et~al.}(2016)\citenamefont {Huntemann}, \citenamefont {Sanner}, \citenamefont {Lipphardt}, \citenamefont {Tamm},\ and\ \citenamefont {Peik}}]{HSL16}%
\BibitemOpen \bibfield  {author} {\bibinfo {author} {\bibfnamefont {N.}~\bibnamefont {Huntemann}}, \bibinfo {author} {\bibfnamefont {C.}~\bibnamefont {Sanner}}, \bibinfo {author} {\bibfnamefont {B.}~\bibnamefont {Lipphardt}}, \bibinfo {author} {\bibfnamefont {Chr.}~\bibnamefont {Tamm}}, \ and\ \bibinfo {author} {\bibfnamefont {E.}~\bibnamefont {Peik}},\ }\href@noop {}
{\bibfield {journal} {\bibinfo  {journal} {Phys. Rev. Lett.}\ }\textbf {\bibinfo {volume} {116}},\ \bibinfo {pages} {063001} (\bibinfo {year} {2016})}\BibitemShut {NoStop}%
\bibitem [{\citenamefont {Kreuter}\ \emph {et~al.}(2005)\citenamefont {Kreuter}, \citenamefont {Becher}, \citenamefont {Lancaster}, \citenamefont {Mundt}, \citenamefont {Russo}, \citenamefont {H{\"a}ffner}, \citenamefont {Roos}, \citenamefont {H{\"a}nsel}, \citenamefont {Schmidt-Kaler}, \citenamefont {Blatt},\ and\ \citenamefont {Safronova}}]{KBL05}%
\BibitemOpen
\bibfield {author} {\bibinfo {author} {\bibfnamefont {A.}~\bibnamefont {Kreuter}}, \bibinfo {author} {\bibfnamefont {C.}~\bibnamefont {Becher}}, \bibinfo {author} {\bibfnamefont {G.~P.~T.}\ \bibnamefont {Lancaster}}, \bibinfo {author} {\bibfnamefont {A.~B.}\ \bibnamefont {Mundt}}, \bibinfo {author} {\bibfnamefont {C.}~\bibnamefont {Russo}}, \bibinfo {author} {\bibfnamefont {H.}~\bibnamefont {H{\"a}ffner}}, \bibinfo {author} {\bibfnamefont {C.}~\bibnamefont {Roos}}, \bibinfo {author} {\bibfnamefont {W.}~\bibnamefont {H{\"a}nsel}}, \bibinfo {author} {\bibfnamefont {F.}~\bibnamefont {Schmidt-Kaler}}, \bibinfo {author} {\bibfnamefont {R.}~\bibnamefont {Blatt}},\ and\ \bibinfo {author} {\bibfnamefont {M.~S.}\ \bibnamefont {Safronova}},\ }\href@noop {}
{\bibfield {journal} {\bibinfo {journal} {Phys. Rev. A}\ }\textbf {\bibinfo {volume} {71}},\ \bibinfo {pages} {032504} (\bibinfo {year} {2005})}\BibitemShut {NoStop}%
\bibitem [{\citenamefont {Gurell}\ \emph {et~al.}(2007)\citenamefont {Gurell}, \citenamefont {Bi{\'e}mont}, \citenamefont {Blagoev}, \citenamefont {Fivet}, \citenamefont {Lundin}, \citenamefont {Mannervik}, \citenamefont {Norlin}, \citenamefont {Quinet}, \citenamefont {Rostohar}, \citenamefont {Royen},\ and\ \citenamefont {Schef}}]{GBB07}%
\BibitemOpen
\bibfield {author} {\bibinfo {author} {\bibfnamefont {J.}~\bibnamefont {Gurell}}, \bibinfo {author} {\bibfnamefont {E.}~\bibnamefont {Bi{\'e}mont}}, \bibinfo {author} {\bibfnamefont {K.}~\bibnamefont {Blagoev}}, \bibinfo {author} {\bibfnamefont {V.}~\bibnamefont {Fivet}}, \bibinfo {author} {\bibfnamefont {P.}~\bibnamefont {Lundin}}, \bibinfo {author} {\bibfnamefont {S.}~\bibnamefont {Mannervik}}, \bibinfo {author} {\bibfnamefont {L.-O.}\ \bibnamefont {Norlin}}, \bibinfo {author} {\bibfnamefont {P.}~\bibnamefont {Quinet}}, \bibinfo {author} {\bibfnamefont {D.}~\bibnamefont {Rostohar}}, \bibinfo {author} {\bibfnamefont {P.}~\bibnamefont {Royen}},\ and\ \bibinfo {author} {\bibfnamefont {P.}~\bibnamefont {Schef}},\ }\href@noop {}
{\bibfield {journal} {\bibinfo {journal} {Phys. Rev. A}\ }\textbf {\bibinfo {volume} {75}},\ \bibinfo {pages} {052506} (\bibinfo {year} {2007})}\BibitemShut {NoStop}%
\bibitem [{\citenamefont {Knoop}\ \emph {et~al.}(1995)\citenamefont {Knoop}, \citenamefont {Vedel},\ and\ \citenamefont {Vedel}}]{KVV95}%
\BibitemOpen
\bibfield {author} {\bibinfo {author} {\bibfnamefont {M.}~\bibnamefont {Knoop}}, \bibinfo {author} {\bibfnamefont {M.}~\bibnamefont {Vedel}},\ and\ \bibinfo {author} {\bibfnamefont {F.}~\bibnamefont {Vedel}},\ }\href@noop {}
{\bibfield {journal} {\bibinfo {journal} {Phys. Rev. A}\ }\textbf {\bibinfo {volume} {52}},\ \bibinfo {pages} {3763} (\bibinfo {year} {1995})}\BibitemShut {NoStop}%
\bibitem [{\citenamefont {Lidberg}\ \emph {et~al.}(1999)\citenamefont {Lidberg}, \citenamefont {Al-Khalili}, \citenamefont {Norlin}, \citenamefont {Royen}, \citenamefont {Tordoir},\ and\ \citenamefont {Mannervik}}]{LAN99}%
\BibitemOpen
\bibfield {author} {\bibinfo {author} {\bibfnamefont {J.}~\bibnamefont {Lidberg}}, \bibinfo {author} {\bibfnamefont {A.}~\bibnamefont {Al-Khalili}}, \bibinfo {author} {\bibfnamefont {L.-O.}\ \bibnamefont {Norlin}}, \bibinfo {author} {\bibfnamefont {P.}~\bibnamefont {Royen}}, \bibinfo {author} {\bibfnamefont {X.}~\bibnamefont {Tordoir}},\ and\ \bibinfo {author} {\bibfnamefont {S.}~\bibnamefont {Mannervik}},\ }\href@noop {}
{\bibfield {journal} {\bibinfo {journal} {J. Phys. B}\ }\textbf {\bibinfo {volume} {32}},\ \bibinfo {pages} {757} (\bibinfo {year} {1999})}\BibitemShut {NoStop}%
\bibitem [{\citenamefont {Barton}\ \emph {et~al.}(2000)\citenamefont {Barton}, \citenamefont {Donald}, \citenamefont {Lucas}, \citenamefont {Stevens}, \citenamefont {Steane},\ and\ \citenamefont {Stacey}}]{BDL00}%
\BibitemOpen
\bibfield {author} {\bibinfo {author} {\bibfnamefont {P.~A.}\ \bibnamefont {Barton}}, \bibinfo {author} {\bibfnamefont {C.~J.~S.}\ \bibnamefont {Donald}}, \bibinfo {author} {\bibfnamefont {D.~M.}\ \bibnamefont {Lucas}}, \bibinfo {author} {\bibfnamefont {D.~A.}\ \bibnamefont {Stevens}}, \bibinfo {author} {\bibfnamefont {A.~M.}\ \bibnamefont {Steane}},\ and\ \bibinfo {author} {\bibfnamefont {D.~N.}\ \bibnamefont {Stacey}},\ }\href@noop {}
{\bibfield {journal} {\bibinfo {journal} {Phys. Rev. A}\ }\textbf {\bibinfo {volume} {62}},\ \bibinfo {pages} {032503} (\bibinfo {year} {2000})}\BibitemShut {NoStop}%
\bibitem [{\citenamefont {Arbes}\ \emph {et~al.}(1994)\citenamefont {Arbes}, \citenamefont {Benzing}, \citenamefont {Gudjons}, \citenamefont {Kurth},\ and\ \citenamefont {Werth}}]{ABG94}%
\BibitemOpen
\bibfield {author} {\bibinfo {author} {\bibfnamefont {F.}~\bibnamefont {Arbes}}, \bibinfo {author} {\bibfnamefont {F.}~\bibnamefont {Benzing}}, \bibinfo {author} {\bibfnamefont {T.}~\bibnamefont {Gudjons}}, \bibinfo {author} {\bibfnamefont {F.}~\bibnamefont {Kurth}},\ and\ \bibinfo {author} {\bibfnamefont {G.}~\bibnamefont {Werth}},\ }\href@noop {}
{\bibfield {journal} {\bibinfo {journal} {Z. Phys. D: At., Mol. Clusters}\ }\textbf {\bibinfo {volume} {29}},\ \bibinfo {pages} {159} (\bibinfo {year} {1994})}\BibitemShut {NoStop}%
\bibitem [{\citenamefont {Block}\ \emph {et~al.}(1999)\citenamefont {Block}, \citenamefont {Rehm}, \citenamefont {Seibert},\ and\ \citenamefont {Werth}}]{BRS99}%
\BibitemOpen
\bibfield {author} {\bibinfo {author} {\bibfnamefont {M.}~\bibnamefont {Block}}, \bibinfo {author} {\bibfnamefont {O.}~\bibnamefont {Rehm}}, \bibinfo {author} {\bibfnamefont {P.}~\bibnamefont {Seibert}},\ and\ \bibinfo {author} {\bibfnamefont {G.}~\bibnamefont {Werth}},\ }\href@noop {}
{\bibfield {journal} {\bibinfo {journal} {Eur. Phys. J. D}\ }\textbf {\bibinfo {volume} {7}},\ \bibinfo {pages} {461} (\bibinfo {year} {1999})}\BibitemShut {NoStop}%
\bibitem [{\citenamefont {Guan}\ \emph {et~al.}(2015)\citenamefont {Guan}, \citenamefont {Huang}, \citenamefont {Liu}, \citenamefont {Bian}, \citenamefont {Shao},\ and\ \citenamefont {Gao}}]{GHL15}%
\BibitemOpen
\bibfield {author} {\bibinfo {author} {\bibfnamefont {H.}~\bibnamefont {Guan}}, \bibinfo {author} {\bibfnamefont {Y.}~\bibnamefont {Huang}}, \bibinfo {author} {\bibfnamefont {P.-L.}\ \bibnamefont {Liu}}, \bibinfo {author} {\bibfnamefont {W.}~\bibnamefont {Bian}}, \bibinfo {author} {\bibfnamefont {H.}~\bibnamefont {Shao}},\ and\ \bibinfo {author} {\bibfnamefont {K.-L.}\ \bibnamefont {Gao}},\ }\href@noop {}
{\bibfield {journal} {\bibinfo {journal} {Chin. Phys. B}\ }\textbf {\bibinfo {volume} {24}},\ \bibinfo {pages} {054213} (\bibinfo {year} {2015})}\BibitemShut {NoStop}%
\bibitem [{\citenamefont {Mannervik}\ \emph {et~al.}(1999)\citenamefont {Mannervik}, \citenamefont {Lidberg}, \citenamefont {Norlin}, \citenamefont {Royen}, \citenamefont {Schmitt}, \citenamefont {Shi},\ and\ \citenamefont {Tordoir}}]{MLN99}%
\BibitemOpen
\bibfield {author} {\bibinfo {author} {\bibfnamefont {S.}~\bibnamefont {Mannervik}}, \bibinfo {author} {\bibfnamefont {J.}~\bibnamefont {Lidberg}}, \bibinfo {author} {\bibfnamefont {L.-O.}\ \bibnamefont {Norlin}}, \bibinfo {author} {\bibfnamefont {P.}~\bibnamefont {Royen}}, \bibinfo {author} {\bibfnamefont {A.}~\bibnamefont {Schmitt}}, \bibinfo {author} {\bibfnamefont {W.}~\bibnamefont {Shi}},\ and\ \bibinfo {author} {\bibfnamefont {X.}~\bibnamefont {Tordoir}},\ }\href@noop {}
{\bibfield {journal} {\bibinfo {journal} {Phys. Rev. Lett.}\ }\textbf {\bibinfo {volume} {83}},\ \bibinfo {pages} {698} (\bibinfo {year} {1999})}\BibitemShut {NoStop}%
\bibitem [{\citenamefont {Bi{\'e}mont}\ \emph {et~al.}(2000)\citenamefont {Bi{\'e}mont}, \citenamefont {Mannervik}, \citenamefont {Norlin}, \citenamefont {Royen}, \citenamefont {Schmitt}, \citenamefont {Shi},\ and\ \citenamefont {Tordoir}}]{BMN00}%
\BibitemOpen
\bibfield {author} {\bibinfo {author} {\bibfnamefont {E.}~\bibnamefont {Bi{\'e}mont}}, \bibinfo {author} {\bibfnamefont {S.}~\bibnamefont {Mannervik}}, \bibinfo {author} {\bibfnamefont {L.-O.}\ \bibnamefont {Norlin}}, \bibinfo {author} {\bibfnamefont {P.}~\bibnamefont {Royen}}, \bibinfo {author} {\bibfnamefont {A.}~\bibnamefont {Schmitt}}, \bibinfo {author} {\bibfnamefont {W.}~\bibnamefont {Shi}},\ and\ \bibinfo {author} {\bibfnamefont {X.}~\bibnamefont {Tordoir}},\ }\href@noop {}
{\bibfield {journal} {\bibinfo {journal} {Eur. Phys. J. D}\ }\textbf {\bibinfo {volume} {11}},\ \bibinfo {pages} {355} (\bibinfo {year} {2000})}\BibitemShut {NoStop}%
\bibitem [{\citenamefont {Gerz}\ \emph {et~al.}(1987)\citenamefont {Gerz}, \citenamefont {Hilberath},\ and\ \citenamefont {Werth}}]{GHW87}%
\BibitemOpen
\bibfield {author} {\bibinfo {author} {\bibfnamefont {C.}~\bibnamefont {Gerz}}, \bibinfo {author} {\bibfnamefont {T.}~\bibnamefont {Hilberath}},\ and\ \bibinfo {author} {\bibfnamefont {G.}~\bibnamefont {Werth}},\ }\href@noop {}
{\bibfield {journal} {\bibinfo {journal} {Z. Phys. D: At., Mol. Clusters}\ }\textbf {\bibinfo {volume} {5}},\ \bibinfo {pages} {97} (\bibinfo {year} {1987})}\BibitemShut {NoStop}%
\bibitem [{\citenamefont {Madej}\ \emph {et~al.}(1990)\citenamefont {Madej}\ and\ \citenamefont {Sankey}}]{MS90}%
\BibitemOpen
\bibfield {author} {\bibinfo {author} {\bibfnamefont {A.~A.}\ \bibnamefont {Madej}}\ and\ \bibinfo {author} {\bibfnamefont {J.~D.}\ \bibnamefont {Sankey}},\ }\href@noop {}
{\bibfield {journal} {\bibinfo {journal} {Opt. Lett.}\ }\textbf {\bibinfo {volume} {15}},\ \bibinfo {pages} {634} (\bibinfo {year} {1990})}\BibitemShut {NoStop}%
\bibitem [{\citenamefont {Letchumanan}\ \emph {et~al.}(2005)\citenamefont {Letchumanan}, \citenamefont {Wilson}, \citenamefont {Gill},\ and\ \citenamefont {Sinclair}}]{LWG05}%
\BibitemOpen
\bibfield {author} {\bibinfo {author} {\bibfnamefont {V.}~\bibnamefont {Letchumanan}}, \bibinfo {author} {\bibfnamefont {M.~A.}\ \bibnamefont {Wilson}}, \bibinfo {author} {\bibfnamefont {P.}~\bibnamefont {Gill}},\ and\ \bibinfo {author} {\bibfnamefont {A.~G.}\ \bibnamefont {Sinclair}},\ }\href@noop {}
{\bibfield {journal} {\bibinfo {journal} {Phys. Rev. A}\ }\textbf {\bibinfo {volume} {72}},\ \bibinfo {pages} {012509} (\bibinfo {year} {2005})}\BibitemShut {NoStop}%
\bibitem [{\citenamefont {Barwood}\ \emph {et~al.}(1993)\citenamefont {Barwood}, \citenamefont {Edwards}, \citenamefont {Gill}, \citenamefont {Klein},\ and\ \citenamefont {Rowley}}]{BEG93}%
\BibitemOpen
\bibfield {author} {\bibinfo {author} {\bibfnamefont {G.~P.}\ \bibnamefont {Barwood}}, \bibinfo {author} {\bibfnamefont {C.~S.}\ \bibnamefont {Edwards}}, \bibinfo {author} {\bibfnamefont {P.}~\bibnamefont {Gill}}, \bibinfo {author} {\bibfnamefont {H.~A.}\ \bibnamefont {Klein}},\ and\ \bibinfo {author} {\bibfnamefont {W.~R.}\ \bibnamefont {Rowley}},\ in\ }\href@noop {}
{\emph {\bibinfo {title} {Eleventh International Conference on Laser Spectroscopy, 1993,}}}\ \bibinfo {note} {edited by \bibfnamefont {L.}~\bibfnamefont {Bloomfield}, \bibfnamefont {T.}~\bibfnamefont {Gallagher} and \bibfnamefont {D.}~\bibfnamefont {Larson}, AIP Conf. Proc.}\ (\bibinfo  {publisher} {AIP, New York},\ \bibinfo {year} {1993}), p. 35 \BibitemShut {NoStop}%
\bibitem [{\citenamefont {Yu}\ \emph {et~al.}(1997)\citenamefont {Yu}, \citenamefont {Nagourney},\ and\ \citenamefont {Dehmelt}}]{YND97}%
\BibitemOpen
\bibfield {author} {\bibinfo {author} {\bibfnamefont {N.}~\bibnamefont {Yu}}, \bibinfo {author} {\bibfnamefont {W.}~\bibnamefont {Nagourney}},\ and\ \bibinfo {author} {\bibfnamefont {H.}~\bibnamefont {Dehmelt}},\ }\href@noop {}
{\bibfield {journal} {\bibinfo {journal} {Phys. Rev. Lett.}\ }\textbf {\bibinfo {volume} {78}},\ \bibinfo {pages} {4898} (\bibinfo {year} {1997})}\BibitemShut {NoStop}%
\bibitem [{\citenamefont {Auchter}\ \emph {et~al.}(2014)\citenamefont {Auchter}, \citenamefont {Noel}, \citenamefont {Hoffman}, \citenamefont {Williams},\ and\ \citenamefont {Blinov}}]{ANH14}%
\BibitemOpen
\bibfield {author} {\bibinfo {author} {\bibfnamefont {C.}~\bibnamefont {Auchter}}, \bibinfo {author} {\bibfnamefont {T.~W.}\ \bibnamefont {Noel}}, \bibinfo {author} {\bibfnamefont {M.~R.}\ \bibnamefont {Hoffman}}, \bibinfo {author} {\bibfnamefont {S.~R.}\ \bibnamefont {Williams}},\ and\ \bibinfo {author} {\bibfnamefont {B.~B.}\ \bibnamefont {Blinov}},\ }\href@noop {}
{\bibfield {journal} {\bibinfo {journal} {Phys. Rev. A}\ }\textbf {\bibinfo {volume} {90}},\ \bibinfo {pages} {060501} (\bibinfo {year} {2014})}\BibitemShut {NoStop}%
\bibitem [{\citenamefont {Nagourney}\ \emph {et~al.}(1986)\citenamefont {Nagourney}, \citenamefont {Sandberg},\ and\ \citenamefont {Dehmelt}}]{NSD86}%
\BibitemOpen
\bibfield {author} {\bibinfo {author} {\bibfnamefont {W.}~\bibnamefont {Nagourney}}, \bibinfo {author} {\bibfnamefont {J.}~\bibnamefont {Sandberg}},\ and\ \bibinfo {author} {\bibfnamefont {H.}~\bibnamefont {Dehmelt}},\ }\href@noop {}
{\bibfield {journal} {\bibinfo {journal} {Phys. Rev. Lett.}\ }\textbf {\bibinfo {volume} {56}},\ \bibinfo {pages} {2797} (\bibinfo {year} {1986})}\BibitemShut {NoStop}%
\bibitem [{\citenamefont {Safronova}\ \emph {et~al.}(2017)\citenamefont {Safronova}, \citenamefont {Safronova},\ and\ \citenamefont {Johnson}}]{SSJ17}%
\BibitemOpen
\bibfield {author} {\bibinfo {author} {\bibfnamefont {U.~I.}\ \bibnamefont {Safronova}}, \bibinfo {author} {\bibfnamefont {M.~S.}\ \bibnamefont {Safronova}},\ and\ \bibinfo {author} {\bibfnamefont {W.~R.}\ \bibnamefont {Johnson}},\ }\href@noop {}
{\bibfield {journal} {\bibinfo {journal} {Phys. Rev. A}\ }\textbf {\bibinfo {volume} {95}},\ \bibinfo {pages} {042507} (\bibinfo {year} {2017})}\BibitemShut {NoStop}%
\bibitem [{\citenamefont {Sahoo}\ \emph {et~al.}(2006)\citenamefont {Sahoo}, \citenamefont {Islam}, \citenamefont {Das}, \citenamefont {Chaudhuri},\ and\ \citenamefont {Mukherjee}}]{SID06}%
\BibitemOpen
\bibfield {author} {\bibinfo {author} {\bibfnamefont {B.~K.}\ \bibnamefont {Sahoo}}, \bibinfo {author} {\bibfnamefont {Md.~R.}\ \bibnamefont {Islam}}, \bibinfo {author} {\bibfnamefont {B.~P.}\ \bibnamefont {Das}}, \bibinfo {author} {\bibfnamefont {R.~K.}\ \bibnamefont {Chaudhuri}},\ and\ \bibinfo {author} {\bibfnamefont {D.}~\bibnamefont {Mukherjee}},\ }\href@noop {}
{\bibfield {journal} {\bibinfo {journal} {Phys. Rev. A}\ }\textbf {\bibinfo {volume} {74}},\ \bibinfo {pages} {062504} (\bibinfo {year} {2006})}\BibitemShut {NoStop}%
\bibitem [{\citenamefont {Vaeck}\ \emph {et~al.}(1992)\citenamefont {Vaeck}, \citenamefont {Godefroid},\ and\ \citenamefont {Froese Fischer}}]{VGF92}%
\BibitemOpen
\bibfield {author} {\bibinfo {author} {\bibfnamefont {N.}~\bibnamefont {Vaeck}}, \bibinfo {author} {\bibfnamefont {M.}~\bibnamefont {Godefroid}},\ and\ \bibinfo {author} {\bibfnamefont {C.}~\bibnamefont {Froese Fischer}},\ }\href@noop {}
{\bibfield {journal} {\bibinfo {journal} {Phys. Rev. A}\ }\textbf {\bibinfo {volume} {46}},\ \bibinfo {pages} {3704} (\bibinfo {year} {1992})}\BibitemShut {NoStop}%
\bibitem [{\citenamefont {Dzuba}\ \emph {et~al.}(2001)\citenamefont {Dzuba}, \citenamefont {Flambaum},\ and\ \citenamefont {Ginges}}]{DFG01}%
\BibitemOpen
\bibfield {author} {\bibinfo {author} {\bibfnamefont {V.~A.}\ \bibnamefont {Dzuba}}, \bibinfo {author} {\bibfnamefont {V.~V.}\ \bibnamefont {Flambaum}},\ and\ \bibinfo {author} {\bibfnamefont {J.~S.~M.}\ \bibnamefont {Ginges}},\ }\href@noop {}
{\bibfield {journal} {\bibinfo {journal} {Phys. Rev. A}\ }\textbf {\bibinfo {volume} {63}},\ \bibinfo {pages} {062101} (\bibinfo {year} {2001})}\BibitemShut {NoStop}%
\bibitem [{\citenamefont {Safronova}\ \emph {et~al.}(2010)\citenamefont {Safronova}, \citenamefont {Johnson},\ and\ \citenamefont {Safronova}}]{SJS10}%
\BibitemOpen
\bibfield {author} {\bibinfo {author} {\bibfnamefont {M.~S.}\ \bibnamefont {Safronova}}, \bibinfo {author} {\bibfnamefont {W.~R.}\ \bibnamefont {Johnson}},\ and\ \bibinfo {author} {\bibfnamefont {U.~I.}\ \bibnamefont {Safronova}},\ }\href@noop {}
{\bibfield {journal} {\bibinfo {journal} {J. Phys. B}\ }\textbf {\bibinfo {volume} {43}},\ \bibinfo {pages} {074014} (\bibinfo {year} {2010})}\BibitemShut {NoStop}%
\bibitem [{\citenamefont {Grant}(2007)}]{Gr07}%
\BibitemOpen
\bibfield {author} {\bibinfo {author} {\bibfnamefont {I.~P.}\ \bibnamefont {Grant}},\ }\href@noop {}
{\emph {\bibinfo {title} {Relativistic Quantum Theory of Atoms and Molecules}}}\ (\bibinfo {publisher} {Springer, New York},\ \bibinfo {year} {2007})\BibitemShut {NoStop}%
\bibitem [{\citenamefont {Safronova}\ \emph {et~al.}(2008)\citenamefont {Safronova}\ and\ \citenamefont {Johnson}}]{SJ08}%
\BibitemOpen
\bibfield {author} {\bibinfo {author} {\bibfnamefont {M.~S.}\ \bibnamefont {Safronova}}\ and\ \bibinfo {author} {\bibfnamefont {W.~R.}\ \bibnamefont {Johnson}},\ }\href@noop {}
{\bibfield {journal} {\bibinfo {journal} {Adv. At. Mol. Opt. Phys.}\ }\textbf {\bibinfo {volume} {55}},\ \bibinfo {pages} {191} (\bibinfo {year} {2008})}\BibitemShut {NoStop}%
\bibitem [{\citenamefont {Froese Fischer}\ \emph {et~al.}(2016)\citenamefont {Froese Fischer}, \citenamefont {Godefroid}, \citenamefont {Brage}, \citenamefont {J\"{o}nsson},\ and\ \citenamefont {Gaigalas}}]{FGB16b}%
\BibitemOpen
\bibfield {author} {\bibinfo {author} {\bibfnamefont {C.}~\bibnamefont {Froese Fischer}}, \bibinfo {author} {\bibfnamefont {M.}~\bibnamefont {Godefroid}}, \bibinfo {author} {\bibfnamefont {T.}~\bibnamefont {Brage}}, \bibinfo {author} {\bibfnamefont {P.}~\bibnamefont {J\"{o}nsson}},\ and\ \bibinfo {author} {\bibfnamefont {G.}~\bibnamefont {Gaigalas}},\ }\href@noop {}
{\bibfield {journal} {\bibinfo {journal} {J. Phys. B}\ }\textbf {\bibinfo {volume} {49}},\ \bibinfo {pages} {182004} (\bibinfo {year} {2016})}\BibitemShut {NoStop}%
\bibitem [{\citenamefont {Mitroy}\ \emph {et~al.}(2008)\citenamefont {Mitroy}, \citenamefont {Zhang},\ and\ \citenamefont {Bromley}}]{MZB08}%
\BibitemOpen
\bibfield {author} {\bibinfo {author} {\bibfnamefont {J.}~\bibnamefont {Mitroy}}, \bibinfo {author} {\bibfnamefont {J.~Y.}\ \bibnamefont {Zhang}},\ and\ \bibinfo {author} {\bibfnamefont {M.~W.~J.}\ \bibnamefont {Bromley}},\ }\href@noop {}
{\bibfield {journal} {\bibinfo {journal} {Phys. Rev. A}\ }\textbf {\bibinfo {volume} {77}},\ \bibinfo {pages} {032512} (\bibinfo {year} {2008})}\BibitemShut {NoStop}%
\bibitem [{\citenamefont {Mitroy}\ \emph {et~al.}(2008)\citenamefont {Mitroy}\ and\ \citenamefont {Zhang}}]{MZ08}%
\BibitemOpen
\bibfield {author} {\bibinfo {author} {\bibfnamefont {J.}~\bibnamefont {Mitroy}}\ and\ \bibinfo {author} {\bibfnamefont {J.~Y.}\ \bibnamefont {Zhang}},\ }\href@noop {}
{\bibfield {journal} {\bibinfo {journal} {Eur. Phys. J. D}\ }\textbf {\bibinfo {volume} {46}},\ \bibinfo {pages} {415} (\bibinfo {year} {2008})}\BibitemShut {NoStop}%
\bibitem [{\citenamefont {Baye}\ and\ \citenamefont {Heenen}(1986)}]{BH86}%
\BibitemOpen
\bibfield {author} {\bibinfo {author} {\bibfnamefont {D.}~\bibnamefont {Baye}}\ and\ \bibinfo {author} {\bibfnamefont {P.-H.}\ \bibnamefont {Heenen}},\ }\href@noop {}
{\bibfield {journal} {\bibinfo {journal} {J. Phys. A}\ }\textbf {\bibinfo {volume} {19}},\ \bibinfo {pages} {2041} (\bibinfo {year} {1986})}\BibitemShut {NoStop}%
\bibitem [{\citenamefont {Baye}(2015)}]{Ba15}%
\BibitemOpen
\bibfield {author} {\bibinfo {author} {\bibfnamefont {D.}~\bibnamefont {Baye}},\ }\href@noop {}
{\bibfield {journal} {\bibinfo {journal} {Phys. Rep.}\ }\textbf {\bibinfo {volume} {565}},\ \bibinfo {pages} {1} (\bibinfo {year} {2015})}\BibitemShut {NoStop}%
\bibitem [{\citenamefont {Vincke}\ \emph {et~al.}(1993)\citenamefont {Vincke}, \citenamefont {Malegat},\ and\ \citenamefont {Baye}}]{VMB93}%
\BibitemOpen
\bibfield {author} {\bibinfo {author} {\bibfnamefont {M.}~\bibnamefont {Vincke}}, \bibinfo {author} {\bibfnamefont {L.}~\bibnamefont {Malegat}},\ and\ \bibinfo {author} {\bibfnamefont {D.}~\bibnamefont {Baye}},\ }\href@noop {}
{\bibfield {journal} {\bibinfo {journal} {J. Phys. B}\ }\textbf {\bibinfo {volume} {26}},\ \bibinfo {pages} {811} (\bibinfo {year} {1993})}\BibitemShut {NoStop}%
\bibitem [{\citenamefont {Baye}\ \emph {et~al.}(2014)\citenamefont {Baye}, \citenamefont {Filippin},\ and\ \citenamefont {Godefroid}}]{BFG14}%
\BibitemOpen
\bibfield {author} {\bibinfo {author} {\bibfnamefont {D.}~\bibnamefont {Baye}}, \bibinfo {author} {\bibfnamefont {L.}~\bibnamefont {Filippin}}, \ and\ \bibinfo {author} {\bibfnamefont {M.}~\bibnamefont {Godefroid}},\ }\href@noop {}
{\bibfield {journal} {\bibinfo {journal} {Phys. Rev. E}\ }\textbf {\bibinfo {volume} {89}},\ \bibinfo {pages} {043305} (\bibinfo {year} {2014})}\BibitemShut {NoStop}%
\bibitem [{\citenamefont {Filippin}\ \emph {et~al.}(2014)\citenamefont {Filippin}, \citenamefont {Godefroid},\ and\ \citenamefont {Baye}}]{FGB14}%
\BibitemOpen
\bibfield {author} {\bibinfo {author} {\bibfnamefont {L.}~\bibnamefont {Filippin}}, \bibinfo {author} {\bibfnamefont {M.}~\bibnamefont {Godefroid}}, \ and\ \bibinfo {author} {\bibfnamefont {D.}~\bibnamefont {Baye}},\ }\href@noop {}
{\bibfield {journal} {\bibinfo {journal} {Phys. Rev. A}\ }\textbf {\bibinfo {volume} {90}},\ \bibinfo {pages} {052520} (\bibinfo {year} {2014})}\BibitemShut {NoStop}%
\bibitem [{\citenamefont {Filippin}\ \emph {et~al.}(2016)\citenamefont {Filippin}, \citenamefont {Godefroid},\ and\ \citenamefont {Baye}}]{FGB16}%
\BibitemOpen
\bibfield {author} {\bibinfo {author} {\bibfnamefont {L.}~\bibnamefont {Filippin}}, \bibinfo {author} {\bibfnamefont {M.}~\bibnamefont {Godefroid}}, \ and\ \bibinfo {author} {\bibfnamefont {D.}~\bibnamefont {Baye}},\ }\href@noop {}
{\bibfield  {journal} {\bibinfo {journal} {Phys. Rev. A}\ }\textbf {\bibinfo {volume} {93}},\ \bibinfo {pages} {012517} (\bibinfo {year} {2016})}\BibitemShut {NoStop}%
\bibitem [{\citenamefont {J\"{o}nsson}\ \emph {et~al.}(2007)\citenamefont {J\"{o}nsson}, \citenamefont {He}, \citenamefont {Froese Fischer},\ and\ \citenamefont {Grant}}]{JHF07}%
\BibitemOpen
\bibfield {author} {\bibinfo {author} {\bibfnamefont {P.}~\bibnamefont {J\"{o}nsson}}, \bibinfo {author} {\bibfnamefont {X.}~\bibnamefont {He}}, \bibinfo {author} {\bibfnamefont {C.}~\bibnamefont {Froese Fischer}},\ and\ \bibinfo {author} {\bibfnamefont {I.~P.}\ \bibnamefont {Grant}},\ }\href@noop {}
{\bibfield {journal} {\bibinfo {journal} {Comput. Phys. Commun.}\ }\textbf {\bibinfo {volume} {177}},\ \bibinfo {pages} {597} (\bibinfo {year} {2007})}\BibitemShut {NoStop}%
\bibitem [{\citenamefont {J\"{o}nsson}\ \emph {et~al.}(2013)\citenamefont {J\"{o}nsson}, \citenamefont {Gaigalas}, \citenamefont {Biero{\'n}}, \citenamefont {Froese Fischer},\ and\ \citenamefont {Grant}}]{JGB13}%
\BibitemOpen
\bibfield {author} {\bibinfo {author} {\bibfnamefont {P.}~\bibnamefont {J\"{o}nsson}}, \bibinfo {author} {\bibfnamefont {G.}~\bibnamefont {Gaigalas}}, \bibinfo {author} {\bibfnamefont {J.}~\bibnamefont {Biero{\'n}}}, \bibinfo {author} {\bibfnamefont {C.}~\bibnamefont {Froese Fischer}},\ and\ \bibinfo {author} {\bibfnamefont {I.~P.}\ \bibnamefont {Grant}},\ }\href@noop {}
{\bibfield {journal} {\bibinfo {journal} {Comput. Phys. Commun.}\ }\textbf {\bibinfo {volume} {184}},\ \bibinfo {pages} {2197} (\bibinfo {year} {2013})}\BibitemShut {NoStop}%
\bibitem [{\citenamefont {Mohr}\ \emph {et~al.}(2016)\citenamefont {Mohr}, \citenamefont {Newell},\ and\ \citenamefont {Taylor}}]{MNT16}%
\BibitemOpen
\bibfield {author} {\bibinfo {author} {\bibfnamefont {P.~J.}\ \bibnamefont {Mohr}}, \bibinfo {author} {\bibfnamefont {D.~B.}\ \bibnamefont {Newell}},\ and\ \bibinfo {author} {\bibfnamefont {B.~N.}\ \bibnamefont {Taylor}},\ }\href@noop {} {\bibfield  {journal} {\bibinfo  {journal} {Rev. Mod. Phys.}\ }\textbf {\bibinfo {volume} {88}},\ \bibinfo {pages} {035009} (\bibinfo {year} {2016})}\BibitemShut {NoStop}%
\bibitem [{\citenamefont {Froese Fischer}\ \emph {et~al.}(1997)\citenamefont {Froese Fischer}, \citenamefont {Brage},\ and\ \citenamefont {J\"{o}nsson}}]{FBJ97}%
\BibitemOpen
\bibfield {author} {\bibinfo {author} {\bibfnamefont {C.}~\bibnamefont {Froese Fischer}}, \bibinfo {author} {\bibfnamefont {T.}~\bibnamefont {Brage}},\ and\ \bibinfo {author} {\bibfnamefont {P.}~\bibnamefont {J\"{o}nsson}},\ }\href@noop {}
{\emph {\bibinfo {title} {Computational Atomic Structure: An MCHF Approach}}}\ (\bibinfo {publisher} {Institute of Physics Publishing, London},\ \bibinfo {year} {1997})\BibitemShut {NoStop}%
\bibitem [{\citenamefont {Zatsarinny}\ \emph {et~al.}(2016)\citenamefont {Zatsarinny}\ and\ \citenamefont {Froese Fischer}}]{ZF16}%
\BibitemOpen
\bibfield {author} {\bibinfo {author} {\bibfnamefont {O.}~\bibnamefont {Zatsarinny}}\ and\ \bibinfo {author} {\bibfnamefont {C.}~\bibnamefont {Froese Fischer}},\ }\href@noop {}
{\bibfield {journal} {\bibinfo {journal} {Comput. Phys. Commun.}\ }\textbf {\bibinfo {volume} {202}},\ \bibinfo {pages} {287} (\bibinfo {year} {2016})}\BibitemShut {NoStop}%
\bibitem [{\citenamefont {Yerokhin}\ \emph {et~al.}(2016)\citenamefont {Yerokhin}, \citenamefont {Buhmann}, \citenamefont {Fritzsche},\ and\ \citenamefont {Surzhykov}}]{YBF16}%
\BibitemOpen
\bibfield {author} {\bibinfo {author} {\bibfnamefont {V.~A.}\ \bibnamefont {Yerokhin}}, \bibinfo {author} {\bibfnamefont {S.~Y.}\ \bibnamefont {Buhmann}}, \bibinfo {author} {\bibfnamefont {S.}~\bibnamefont {Fritzsche}}, \ and\ \bibinfo {author} {\bibfnamefont {A.}~\bibnamefont {Surzhykov}},\ }\href@noop {}
{\bibfield  {journal} {\bibinfo {journal} {Phys. Rev. A}\ }\textbf {\bibinfo {volume} {94}},\ \bibinfo {pages} {032503} (\bibinfo {year} {2016})}\BibitemShut {NoStop}%
\bibitem [{\citenamefont {Hameed}\ \emph {et~al.}(1968)\citenamefont {Hameed}, \citenamefont {Herzenberg},\ and\ \citenamefont {James}}]{HHJ68}%
\BibitemOpen
\bibfield  {author} {\bibinfo {author} {\bibfnamefont {S.}~\bibnamefont {Hameed}}, \bibinfo {author} {\bibfnamefont {A.}~\bibnamefont {Herzenberg}}, \ and\ \bibinfo {author} {\bibfnamefont {M.~G.}\ \bibnamefont {James}},\ }\href@noop {}
{\bibfield {journal} {\bibinfo {journal} {J. Phys. B}\ }\textbf {\bibinfo {volume} {1}},\ \bibinfo {pages} {822} (\bibinfo {year} {1968})}\BibitemShut {NoStop}%
\bibitem [{\citenamefont {Norcross}\ and\ \citenamefont {Seaton}(1976)}]{NS76}%
\BibitemOpen
\bibfield {author} {\bibinfo {author} {\bibfnamefont {D.~W.}\ \bibnamefont {Norcross}}\ and\ \bibinfo {author} {\bibfnamefont {M.~J.}\ \bibnamefont {Seaton}},\ }\href@noop {}
{\bibfield {journal} {\bibinfo {journal} {J. Phys. B}\ } \textbf {\bibinfo {volume} {9}},\ \bibinfo {pages} {2983} (\bibinfo {year} {1976})}\BibitemShut {NoStop}%
\bibitem [{\citenamefont {Hibbert}(1982)}]{Hi82}%
\BibitemOpen
\bibfield {author} {\bibinfo {author} {\bibfnamefont {A.}~\bibnamefont {Hibbert}},\ }\href@noop {}
{\bibfield {journal} {\bibinfo {journal} {Adv. At. Mol. Phys.}\ }\textbf {\bibinfo {volume} {18}},\ \bibinfo {pages} {309} (\bibinfo {year} {1982})}\BibitemShut {NoStop}%
\bibitem [{\citenamefont {Mitroy}\ and\ \citenamefont {Norcross}(1988)}]{MN88}%
\BibitemOpen
\bibfield {author} {\bibinfo {author} {\bibfnamefont {J.}~\bibnamefont {Mitroy}}\ and\ \bibinfo {author} {\bibfnamefont {D.~W.}\ \bibnamefont {Norcross}},\ }\href@noop {}
{\bibfield {journal} {\bibinfo {journal} {Phys. Rev. A}\ } \textbf {\bibinfo {volume} {37}},\ \bibinfo {pages} {3755} (\bibinfo {year} {1988})}\BibitemShut {NoStop}%
\bibitem [{\citenamefont {Hameed}(1972)}]{Ha72}%
\BibitemOpen
\bibfield {author} {\bibinfo {author} {\bibfnamefont {S.}~\bibnamefont {Hameed}},\ }\href@noop {}
{\bibfield {journal} {\bibinfo {journal} {J. Phys. B}\ }\textbf {\bibinfo {volume} {5}},\ \bibinfo {pages} {746} (\bibinfo {year} {1972})}\BibitemShut {NoStop}%
\bibitem [{\citenamefont {Goldman}\ and\ \citenamefont {Drake}(1981)}]{GD81}%
\BibitemOpen
\bibfield {author} {\bibinfo {author} {\bibfnamefont {S.~P.}\ \bibnamefont {Goldman}}\ and\ \bibinfo {author} {\bibfnamefont {G.~W.~F.}\ \bibnamefont {Drake}},\ }\href@noop {}
{\bibfield {journal} {\bibinfo {journal} {Phys. Rev. A}\ } \textbf {\bibinfo {volume} {24}},\ \bibinfo {pages} {183} (\bibinfo {year} {1981})}\BibitemShut {NoStop}%
\bibitem [{\citenamefont {Grant}(1974)\citenamefont {Grant}}]{Gr74}%
\BibitemOpen
\bibfield {author}{ \bibinfo {author} {\bibfnamefont {I.~P.}\ \bibnamefont {Grant}},\ }\href@noop {}
\bibfield {journal}{\bibinfo {journal} {J. Phys. B}\ } \textbf {\bibinfo {volume} {7}},\ \bibinfo {pages} {1458} (\bibinfo {year} {1974})\BibitemShut {NoStop}%
\bibitem [{\citenamefont {Santos}\ \emph {et~al.}(1998)\citenamefont {Santos}, \citenamefont {Parente},\ and\ \citenamefont {Indelicato}}]{SPI98}%
\BibitemOpen
\bibfield {author} {\bibinfo {author} {\bibfnamefont {J.~P.}\ \bibnamefont {Santos}}, \bibinfo {author} {\bibfnamefont {F.}~\bibnamefont {Parente}}, \ and\ \bibinfo {author} {\bibfnamefont {P.}~\bibnamefont {Indelicato}},\ }\href@noop {}
{\bibfield {journal} {\bibinfo {journal} {Eur. Phys. J. D}\ } \textbf {\bibinfo {volume} {3}},\ \bibinfo {pages} {43} (\bibinfo {year} {1998})}\BibitemShut {NoStop}%
\bibitem [{\citenamefont {Guet}\ and\ \citenamefont {Johnson}(2007)}]{GJ07}%
\BibitemOpen
\bibfield {author} {\bibinfo {author} {\bibfnamefont {C.}~\bibnamefont {Guet}}\ and\ \bibinfo {author} {\bibfnamefont {W.~R.}\ \bibnamefont {Johnson}},\ }\href@noop {}
{\bibfield {journal} {\bibinfo {journal} {Phys. Rev. A}\ } \textbf {\bibinfo {volume} {76}},\ \bibinfo {pages} {039905(E)} (\bibinfo {year} {2007})}\BibitemShut {NoStop}%
\bibitem [{\citenamefont {Abramowitz}\ and\ \citenamefont {Stegun}(1965)}]{AS65}%
\BibitemOpen
\bibfield {author} {\bibinfo {author} {\bibfnamefont {M.}~\bibnamefont {Abramowitz}}\ and\ \bibinfo {author} {\bibfnamefont {I.~A.}\ \bibnamefont {Stegun}},\ }\href@noop {}
{\emph {\bibinfo {title} {Handbook of Mathematical Functions}}}\ (\bibinfo  {publisher} {Dover, New York}, \bibinfo {year} {1965})\BibitemShut {NoStop}%
\bibitem [{\citenamefont {Szeg\"o}(1967)}]{Sz67}%
\BibitemOpen
\bibfield {author} {\bibinfo {author} {\bibfnamefont {G.}~\bibnamefont {Szeg\"o}},\ }\href@noop {}
{\emph {\bibinfo {title} {Orthogonal polynomials}}}\ (\bibinfo  {publisher} {Am. Math. Soc, Providence, RI},\ \bibinfo {year} {1967})\BibitemShut {NoStop}%
\bibitem [{\citenamefont {Baye}(1995)}]{Ba95}%
\BibitemOpen
\bibfield {author} {\bibinfo {author} {\bibfnamefont {D.}~\bibnamefont {Baye}},\ }\href@noop {}
{\bibfield {journal} {\bibinfo {journal} {J. Phys. B}\ }\textbf {\bibinfo {volume} {28}},\ \bibinfo {pages} {4399} (\bibinfo {year} {1995})}\BibitemShut {NoStop}%
\bibitem [{\citenamefont {Baye}\ \emph {et~al.}(2002)\citenamefont {Baye}, \citenamefont {Hesse},\ and\ \citenamefont {Vincke}}]{BHV02}%
\BibitemOpen
\bibfield {author} {\bibinfo {author} {\bibfnamefont {D.}~\bibnamefont {Baye}}, \bibinfo {author} {\bibfnamefont {M.}~\bibnamefont {Hesse}}, \ and\ \bibinfo {author} {\bibfnamefont {M.}~\bibnamefont {Vincke}},\ }\href@noop {}
{\bibfield {journal} {\bibinfo {journal} {Phys. Rev. E}\ }\textbf {\bibinfo {volume} {65}},\ \bibinfo {pages} {026701} (\bibinfo {year} {2002})}\BibitemShut {NoStop}%
\bibitem [{\citenamefont {Hartree}(1957)}]{Ha57}%
\BibitemOpen
\bibfield {author} {\bibinfo {author} {\bibfnamefont {D.~R.}\ \bibnamefont {Hartree}},\ }\href@noop {}
{\emph {\bibinfo {title} {The calculation of Atomic Structures}}}\ (\bibinfo {publisher} {John Wiley \& Sons, New York},\ \bibinfo {year} {1957})\BibitemShut {NoStop}%
\bibitem [{\citenamefont {Kramida}\ \emph {et~al.}(2015)\citenamefont {Kramida}, \citenamefont {Ralchenko}, \citenamefont {Reader},\ and\ \citenamefont {NIST ASD Team}}]{KRR15}%
\BibitemOpen
\bibfield {author} {\bibinfo {author} {\bibfnamefont {A.}~\bibnamefont {Kramida}}, \bibinfo {author} {\bibfnamefont {Y.}~\bibnamefont {Ralchenko}}, \bibinfo {author} {\bibfnamefont {J.}~\bibnamefont {Reader}},\ and\ \bibinfo {author} {\bibnamefont {NIST ASD Team}},\ }\href@noop {}
{\bibinfo {title} {NIST Atomic Spectra Database (version 5.4)}}\ (\bibinfo {year} {2016}), \bibinfo {note} {\url{http://physics.nist.gov/asd}}\BibitemShut {NoStop}%
\bibitem [{\citenamefont {Johnson}\ \emph {et~al.}(1983)\citenamefont {Johnson}, \citenamefont {Kolb},\ and\ \citenamefont {Huang}}]{JKH83}%
\BibitemOpen
\bibfield {author} {\bibinfo {author} {\bibfnamefont {W.~R.}\ \bibnamefont {Johnson}}, \bibinfo {author} {\bibfnamefont {D.}~\bibnamefont {Kolb}},\ and\ \bibinfo {author} {\bibfnamefont {K.~N.}\ \bibnamefont {Huang}},\ }\href@noop {}
{\bibfield {journal} {\bibinfo {journal} {At. Data Nucl. Data Tables}\ }\textbf {\bibinfo {volume} {28}},\ \bibinfo {pages} {333} (\bibinfo {year} {1983})}\BibitemShut {NoStop}%
\bibitem [{\citenamefont {Chattopadhyay}\ \emph {et~al.}(1983)\citenamefont {Chattopadhyay}, \citenamefont {Mani},\ and\ \citenamefont {Angom}}]{CMA13}%
\BibitemOpen
\bibfield {author} {\bibinfo {author} {\bibfnamefont {S.}~\bibnamefont {Chattopadhyay}}, \bibinfo {author} {\bibfnamefont {B.~K.}\ \bibnamefont {Mani}},\ and\ \bibinfo {author} {\bibfnamefont {D.}~\bibnamefont {Angom}},\ }\href@noop {}
{\bibfield {journal} {\bibinfo {journal} {Phys. Rev. A}\ }\textbf {\bibinfo {volume} {87}},\ \bibinfo {pages} {062504} (\bibinfo {year} {2013})}\BibitemShut {NoStop}%
\bibitem [{\citenamefont {Chang}(1983)}]{Ch83}%
\BibitemOpen
\bibfield {author} {\bibinfo {author} {\bibfnamefont {E.~S.}\ \bibnamefont {Chang}},\ }\href@noop {}
{\bibfield {journal} {\bibinfo {journal} {J. Phys. B: At. Mol. Phys.}\ }\textbf {\bibinfo {volume} {16}},\ \bibinfo {pages} {L539} (\bibinfo {year} {1983})}\BibitemShut {NoStop}%
\bibitem [{\citenamefont {Nunkaew}\ \emph {et~al.}(2009)\citenamefont {Nunkaew}, \citenamefont {Shuman},\ and\ \citenamefont {Gallagher}}]{NSG09}%
\BibitemOpen
\bibfield {author} {\bibinfo {author} {\bibfnamefont {J.}~\bibnamefont {Nunkaew}}, \bibinfo {author} {\bibfnamefont {E.~S.}\ \bibnamefont {Shuman}},\ and\ \bibinfo {author} {\bibfnamefont {T.~F.}\ \bibnamefont {Gallagher}},\ }\href@noop {}
{\bibfield {journal} {\bibinfo {journal} {Phys. Rev. A}\ }\textbf {\bibinfo {volume} {79}},\ \bibinfo {pages} {054501} (\bibinfo {year} {2009})}\BibitemShut {NoStop}%
\bibitem [{\citenamefont {Snow}\ \emph {et~al.}(2007)\citenamefont {Snow}\ and\ \citenamefont {Lundeen}}]{SL07}%
\BibitemOpen
\bibfield {author} {\bibinfo {author} {\bibfnamefont {E.~L.}\ \bibnamefont {Snow}}\ and\ \bibinfo {author} {\bibfnamefont {S.~R.}\ \bibnamefont {Lundeen}},\ }\href@noop {}
{\bibfield {journal} {\bibinfo {journal} {Phys. Rev. A}\ }\textbf {\bibinfo {volume} {76}},\ \bibinfo {pages} {052505} (\bibinfo {year} {2007})}\BibitemShut {NoStop}%
\bibitem [{\citenamefont {Safronova}\ \emph {et~al.}(2010)\citenamefont {Safronova}, \citenamefont {Johnson},\ and\ \citenamefont {Safronova}}]{SJS17}%
\BibitemOpen
\bibfield {author} {\bibinfo {author} {\bibfnamefont {M.~S.}\ \bibnamefont {Safronova}}, \bibinfo {author} {\bibfnamefont {W.~R.}\ \bibnamefont {Johnson}},\ and\ \bibinfo {author} {\bibfnamefont {U.~I.}\ \bibnamefont {Safronova}},\ }\href@noop {}
{\bibfield {journal} {\bibinfo {journal} {J. Phys. B}\ }\textbf {\bibinfo {volume} {50}},\ \bibinfo {pages} {189501} (\bibinfo {year} {2017})}\BibitemShut {NoStop}%
\end{thebibliography}
\end{document}